\documentclass[final,authoryear,5p,times,twocolumn]{elsarticle}

\newcommand{\dev}{\ensuremath{\,\mathrm{d}}}

\usepackage{amsmath}

\title{Simulation and physical model based gamma-ray burst afterglow analysis}

\author[mpe]{H.J.~van Eerten\fnref{fn1}}
\ead{hveerten@mpe.mpg.de}

\address[mpe]{Max-Plank-institut f\"ur Extraterrestrische Physik, Giessenbachstra\ss e 1, D-85748 Garching, Germany}

\fntext[fn1]{Alexander von Humboldt Fellow.}

\begin{document}

\begin{abstract}
Advances in our numerical and theoretical understanding of gamma-ray burst afterglow processes allow us to construct models capable of dealing with complex relativistic jet dynamics and non-thermal emission, that can be compared directly to data from instruments such as Swift. Because afterglow blast waves and power law spectra are intrinsically scale-invariant under changes of explosion energy and medium density, templates can be generated from large-scale hydrodynamics simulations. This allows for iterative template-based model fitting using the physical model parameters (quantifying the properties of the burster, emission and observer) directly as fit variables. Here I review how such an approach to afterglow analysis works in practice, paying special attention to the underlying model assumptions, possibilities, caveats and limitations of this type of analysis. Because some model parameters can be degenerate in certain regions of parameter space, or unconstrained if data in a limited number of a bands is available, a Bayesian approach is a natural fit. The main features of the standard afterglow model are reviewed in detail.
\end{abstract}

\maketitle

\section{Introduction}

Gamma-ray bursts (GRBs) are the most luminous explosions occurring in the universe and a key target for many active and upcoming rapid cadence survey programs from radio to gamma-rays. We know they most likely involve the merging of neutron stars (making them prime gravitational wave counterpart candidates) and massive star collapse, and that some form of relativistic outflow is launched. The merger scenario \citep{Eichler1989, Paczynski1991} has been tied to short GRBs, while long GRBs are securely connected to massive star collapse \citep{Woosley1993, Paczynski1998, MacFadyenWoosley1999} through observations of coincident supernovae \citep{Galama1998}.  The separation between `short' and `long' burst durations lies around 2 s. \citep{Kouveliotou1993}.

 We also know this blast wave eventually generates an afterglow from X-rays to radio, as was predicted originally by way of the \emph{fireball} model which explained GRB prompt emission in terms of optically thin synchrotron emission from colliding shells within a hydrodynamically launched relativistic outflow \citep{ReesMeszaros1992, MeszarosRees1997}. During the afterglow stage the emission is dominated by the forward shock interacting with the external medium, with electrons being accelerated to relativistic velocities at the shock front and interacting with the locally generated magnetic field in order to produce synchrotron emission. Afterglow emission is not unique to the synchrotron internal shock model, and a decelerating afterglow-stage blast wave can be associated with a number of mechanisms for jet launching (such as magnetically dominated jets, e.g. \citealt{Drenkhahn2002}) and/or prompt emission (such as photospheric emission models, e.g. \citealt{MeszarosRees2000}).

Afterglow observations have proven extremely useful for a number of reasons and have been instrumental in establishing the extra-galactic nature of GRB \citep{Costa1997, vanParadijs1997}. They provide insight in the environment of the burster (constraining circumburst medium density, amount of dust-extinction), the physical properties of the progenitor (via explosion energy), the physics of jet launching (via jet collimation angle), and the fundamental plasma kinetic theory of relativistic shocks (via micro-physical parameters describing magnetic field generation and electron acceleration). Most of this is done through interpreting the evolution of the characteristics of the synchrotron spectrum in terms of flux equations derived from dynamical blast wave models in the self-similar ultra-relativistic stage \citep{Blandford1976}, the late self-similar non-relativistic stage \citep{Taylor1950, Sedov1959}, or, in a more recent development, from complex multi-dimensional, trans-relativistic evolution in-between. 

In this review, I describe the basic aspects of afterglow models, and the challenges involved in performing data analysis based directly on physical models (as opposed to post-hoc interpretation of purely heuristic functions such as power laws, that capture the shape of the data of light curves and spectra in a simplified manner). In section \ref{dynamics_section}, the dynamics of blast waves from ultra-relativistic to non-relativistic are discussed and self-contained models are provided. Afterglow emission is discussed in section \ref{emission_section} and it is reviewed how flux equations for afterglow spectra are derived. Hydrodynamical simulations and model-based afterglow fitting are discussed in section \ref{fitting_section}.

\section{The dynamics of afterglow blast waves}
\label{dynamics_section}

\subsection{Simple approximations for spherical flow}

Under the assumption of spherical flow, the radius $R$ of the afterglow blast wave is fixed from conservation of explosion energy $E$ in the blast wave (assuming adiabatic expansion). The conditions at the front of the shock are set by the shock-jump conditions. There exists a reasonable approximation to the ideal gas equation of state (EOS) that covers the transition between relativistic and non-relativistic temperatures \citep{Mignone2005}:
\begin{equation}
p / (\rho c^2) = \frac{e / (\rho c^2)}{3} \frac{2 + e / (\rho c^2)}{1 + e / (\rho c^2)},
\end{equation}
where $p$ pressure, $c$ speed of light, $\rho$ co-moving density, $e$ internal energy density excluding rest mass. Using this EOS, the shock-jump conditions can be simplified to
\begin{eqnarray}
\rho & = & 4 \gamma \rho_{ext,} \nonumber \\
\quad e & = & 4 \gamma(\gamma - 1) \rho_{ext} c^2, \nonumber \\
\quad p & = & \frac{4}{3} (\gamma^2 - 1) \rho_{ext} c^2, \nonumber \\
\quad \Gamma^2 & = & \frac{(4 \gamma^2 - 1)^2}{8 \gamma^2 + 1},
\label{shock_jump_equation}
\end{eqnarray}
valid throughout the evolution of the blast wave (and where $\gamma$ the fluid Lorentz factor at the shock front, $\rho_{ext}$ the mass density in front of the shock and $\Gamma$ the Lorentz factor of the shock). The upstream density can be allowed to depend on radius, i.e $\rho_{ext} \equiv \rho_{ref} (R / R_{ref})^{-k}$, with $k = (0, 2)$ covering respectively a homogeneous interstellar-medium (ISM) type and a stellar-wind type environment. The concise description offered by eq. \ref{shock_jump_equation} was pointed out previously by \cite{Uhm2011, vanEerten2013}, and this particular EOS has been used in numerical \citep{ZhangMacFadyen2009, vanEerten2010transrelativistic, vanEerten2011chromaticbreak} and theoretical \citep{Nava2013} analysis. Although in the non-relativistic limit the jump in density is fixed (depending on the polytropic index; the jump of 4 from the equations above applies to an ideal gas), the shock-jump conditions indicate a special feature of relativistic flows, where the density jump can become arbitrarily high. This effect is even stronger when expressed in the lab frame, where an additional factor of $\gamma$ applies.

Further assuming a homogeneous shell model, the radius of the blast wave can be expressed exactly as \citep{vanEerten2013}:
\begin{equation}
E / (M c^2) = \beta^2 (4 \gamma^2 - 1) / 3,
\label{shell_model_equation}
\end{equation}
where $M$ the swept-up mass (i.e. a proxy for radius) and $\beta$ the fluid velocity $v$ in units of $c$. The width of the shell will always be $\Delta R = R / (12 \gamma^2)$. This can be established by taking the downstream density according to the shock-jump conditions and, under the assumption of a homogeneous shell, moving down in radius until all swept-up matter is accounted for.  Due to its inclusion of a simplified EOS, the blast wave model described by eq. \ref{shell_model_equation} is about the most concise analytically tractable approximation possible. 

Other simplified trans-relativistic spherical shell models exist in the literature (e.g. \citealt{Piran1999, HuangDaiLu1999}, who omit pressure, or \citealt{Peer2012}, who includes pressure); in practice, the differences between adiabatic expansion models are very minor, as long as the same asymptotes are reproduced (\citealt{Piran1999} leads to a non-relativistic asymptote different from the self-similar asymptote). A more pronounced difference between possible shell evolutions is that between radiative and adiabatic expansion, and between large and small initial mass content (the latter discussed separately in section \ref{reverse_shock_section}). In the radiative case, blast wave Lorentz factor and radius are still dictated by the energy within the blast wave, only now this energy is diminishing noticeably due to radiative losses, leading to faster deceleration. Calculating the cumulative energy loss from standard synchrotron afterglow emission (as covered below in section \ref{emission_section}) for a shell model will typically only add up to only a few percent well into the non-relativistic stage, justifying the adiabatic assumption from the beginning of this section. However, under certain circumstances, for example when the unshocked medium contains many electron-positron pairs triggered by prompt emission photons \citep{ThompsonMadau2000, Beloborodov2002}, conditions leading to radiative blast waves may arise at early times. Observational support for initially radiative blast waves is offered by Fermi LAT gamma-ray detections, when interpreted in the afterglow blast wave framework (see e.g. \citealt{Ghisellini2010, Nava2013}). Most of the cited shell models incorporate the possibility of a significant radiative energy loss term.

\subsection{The non-relativistic Sedov-Taylor-von Neumann self-similar solution}

A full solution for the blast wave in the radial flow case requires equations for the fluid profile everywhere in the flow, not just behind the shock front. For the limiting cases, when $\beta \uparrow 1$ or $\beta \downarrow 0$, this solution is provided by the aforementioned self-similar solutions, since dimensional analysis indicates only one possible dimensionless combination for the remaining variables (i.e. $E$, $\rho_{ext}$, radius $r$, lab frame time $t$). These analytical solutions tend to be unwieldy, but sometimes reduce to very simple form. An example of the latter is the non-relativistic stellar wind ($\rho_{ext} \propto r^{-2}$) case, where
\begin{equation}
\rho = 4 (r / R) \rho_{ext}, \quad v =  r / (2t), \quad p = \rho_{ext} r^3 / (3 R t^2).
\label{ST_wind_solution_equation}
\end{equation}
Here $r / R$ plays the role of the self-similarity variable, and we have used an ideal gas with polytropic index $\hat{\gamma} = 5/3$. Clearly, these reduce to the non-relativistic shock-jump conditions when $r \to R$ and $\gamma^2 \downarrow 1 + \beta^2$, as can be seen from eq. \ref{shock_jump_equation}. The radius $R$ is in this case given by $R = (12 \pi / 50 )^{1/3} (E t^2 / [\rho_{ref} R_{ref}^k])^{1/3}$. The mass $\dev M$ within a shell at radius $r$ is $16 \pi (r^3 / R) \rho_{ext} \dev r$, confirming once more that the swept-up mass in the blast wave is concentrated at the front.

The full solution for arbitrary $k$ can be obtained e.g. by generalizing the $k=0$ case as described in \cite{LandauLifshitz1959}:
\begin{eqnarray}
v & \equiv & \frac{2}{5-k} \frac{r}{t} V(\xi), \nonumber \\
\rho & \equiv & \rho_{ext} G(\xi), \nonumber \\
c_s^2 & \equiv & \frac{4r^2}{(5-k)^2 t^2} Z(\xi).
\label{ST_fluid_equation}
\end{eqnarray}
Here $c_s$ is the speed of sound. In our case, $p = 3 \rho c_s^2 / 5$. The self-similarity variable $\xi \equiv r / R$ is equal to 1 at the shock front, and the shock-jump conditions therefore yield $V(1) = 3/4$, $G(1) = 4$, $Z(1) = 5/16$ as boundary conditions for the self-similar functions. Plugging the self-similar Ansatz provided by eq. \ref{ST_fluid_equation} into the fluid hydrodynamical equations and solving the resulting differential equations, eventually yields (excluding $k = 2$, which is a special case where terms drop out of the equations early on):
\begin{eqnarray}
\xi^{5-k} & = & \left( \frac{V}{V(1)} \right)^{-2} \left( \frac{-5+k+4V}{-5+k+4V(1)} \right)^{\nu_1} \left( \frac{5V-3}{5 V(1) - 3} \right)^{\nu_2}, \nonumber
\end{eqnarray}
\begin{equation*}
Z = \frac{5(1-V)V^2}{3(5V-3)},
\end{equation*}
\begin{eqnarray}
G & = & 4 \left( \frac{5V-3}{5V(1)-3} \right)^{\nu_3} \left( \frac{4V-5+k}{4V(1)-5+k} \right)^{\nu_4} \left( \frac{V-1}{V(1)-1} \right)^{\nu_5} \nonumber \\ 
 & & \times \left( \frac{V}{V(1)} \right)^{\nu_6}, \nonumber
\end{eqnarray}
\begin{equation}
\nu_1 = \frac{2(41-26k+5k^2)}{3(5k-13)}, \nonumber
\end{equation}
\begin{equation}
\nu_2 = \frac{2(k - 5)}{5k-13}, \qquad \nu_3 = \frac{5k - 9}{5k - 13}, \nonumber 
\end{equation}
\begin{equation}
\nu_4 = \frac{2(7k-15)(5k^2-26k+41)}{3(k-1)(k-5)(5k-13)}, \nonumber
\end{equation}
\begin{equation}
\nu_5 = - \frac{2(9-4k)}{3(k-1)}, \qquad \nu_6 = -\frac{2k}{k - 5}.
\end{equation}

From dimensional analysis, we obtain for the radius
\begin{equation}
R = \hat{\beta} \left( \frac{E t^2}{\rho_{ref} R_{ref}^k} \right)^{1/(5-k)},
\end{equation}
where $\hat{\beta}$ can be found using energy conservation, leading to
\begin{equation}
\hat{\beta}^{k-5} = \frac{16 \pi}{(5-k)^2} \int_0^1 G [ \frac{1}{2} V^2 + \frac{9}{10}Z ] \xi^4 \dev \xi.
\label{ST_energy_equation}
\end{equation}
In the ISM case, $\hat{\beta} \approx 1.15$, in the stellar-wind case, $\hat{\beta} \approx 0.92$.

The late-time non-relativistic solution primary applies to radio observations (e.g. \citealt{WaxmanKulkarniFrail1998}), since emission in higher frequency bands such as optical and X-rays will have already dropped below the detection threshold of most instruments. The distance scale at which a typical blast wave becomes non-relativistic, is vast (as can be confirmed by plugging $\beta \gamma = 1$ into equation \ref{shell_model_equation}). The Sedov-Taylor solution for nonzero $k$ provided above, is therefore not likely to occur in nature in such a clean fashion. The blast wave has expanded well beyond the sphere of influence of its progenitor, making it more likely that its current environment is approximately ISM-like, or shaped by some complex interaction between wind bubbles from surrounding stars \citep{MimicaGiannios2011}. Nevertheless, the non-zero $k$ case is relevant as an asymptotic solution to long-term evolution of non-ISM hydrodynamical blast wave simulations (e.g. those done by \citealt{DeColle2012}).

\subsection{The ultra-relativistic Blandford-McKee self-similar solution}

On the other end of the velocity spectrum sits the Blandford-McKee self-similar solution for ultra-relativistic flow \citep{Blandford1976}. As already suggested by the facts that in this stage $\beta = 1 - 1 / (2 \gamma^2) + O(\gamma^{-4})$, and the width of the shell $\Delta R \propto R / \gamma^2$, a Taylor-series expansion around $\gamma^{-1} \downarrow 0$ for the fluid profile will typically have $\gamma^{-2}$, rather than $\gamma^{-1}$ as its first non-constant contributing order. From energy conservation within the blast wave that moves at nearly the speed of light, one obtains:
\begin{equation}
E = \frac{8 \pi \rho_{ref} R_{ref}^k c^{5-k} \Gamma^2 t^{3-k}}{17 - 4k},
\label{BM_energy_equation}
\end{equation}
where $\Gamma$ the shock Lorentz factor. The numerical factor in this equation follows again from integrating the (rest frame) energy density over the self-similar fluid profile (provided below), as in eq. \ref{ST_energy_equation}. Note that the shocked fluid is relativistically hot (as can be seen from the shock-jump conditions in eq. \ref{shock_jump_equation}), and $\hat{\gamma} = 4/3$. Since according to eq. \ref{BM_energy_equation}, $\Gamma^2 \propto t^{k-3}$, it also follows that
\begin{equation}
R = c t \left( 1 - \frac{1}{2(4-k)\Gamma^2} \right).
\end{equation}
This equation explains the extreme variability of GRBs and early afterglow. A photon emitted from that part of the shock front that is moving directly towards us, is observed at
\begin{equation}
\frac{t_{\oplus}}{(1+z)} = ct - R = \frac{t}{2(4-k)\Gamma^2},
\label{obs_time_equation}
\end{equation}
where $z$ gravitational redshift, and if the observer time is set to zero at the point when the GRB is first observed. The Lorentz factors can reach incredibly high values, 100-1000 and beyond (see e.g. \citealt{Racusin2011}).

Continuing the Blandford-McKee solution, we again combine the shock-jump conditions and hydrodynamical equations with the self-similarity Ansatz, this time for self-similarity variable $\chi$, and obtain
\begin{eqnarray}
p & = & \frac{2}{3} \rho_{ext} c^2 \Gamma^2 \chi^{-(17-4k)/(12-3k)}, \nonumber \\
\gamma^2 & = & \frac{1}{2} \Gamma^2 \chi^{-1}, \nonumber \\
\rho' & = & 2 \rho_{ext} \Gamma^2 \chi^{-(7-2k)/(4-k)}, \nonumber \\
\chi & = & [1 + 2 (4 - k) \Gamma^2 ] (1 - r/[ct]),
\label{BM_equation}
\end{eqnarray}
where $\rho'$ expressed in the lab frame. The ultra-relativistic limits of eq. \ref{shock_jump_equation} are reproduced taking $\chi \downarrow 1$, corresponding to the position of the shock front. In contrast to the non-relativistic self-similar solution, the relativistic version does not work all the way to the origin, but applies only to the relativistic part of the outflow, which is where almost all matter and energy reside anyway. In numerical simulations, non-physical values can be avoided by simply adding 1 to the profile for $\gamma^2$, when setting up the Blandford-McKee solution as initial conditions.

\subsection{Jetted outflow}

A major complication to the simplified picture sketched above, is that GRB blast waves are extremely likely to be collimated into two diametrically opposite jets with narrow opening angles $\theta_0$ \citep{Rhoads1997, Rhoads1999}. Due to strong relativistic beaming, this is not immediately apparent in the light curve, as only a small patch $\dev \theta \sim 1 / \gamma$ is visible at first. Unless the geometry of the outflow is radically different from radial flow (e.g. cylindrical, see \citealt{ChengHuangLu2001}) or the fluid properties are strongly dependent on angle (e.g. as in `structured jet' models, \citealt{Rossi2002}), it is not possible to tell apart collimated and spherical flow at this point. Additionally, if the outflow was launched radially, it will take time for causal contact to be established across angles and for sideways motion to become apparent in the observer frame, postponing a deviation from radial flow even for a conic wedge of limited opening angle. Sky images are of no help either, since GRBs are typically too distant for spatially resolved observations (the exception being GRB 030329, \citealt{Taylor2004, Oren2004}).

Jet collimation therefore has to be inferred indirectly, and this can be done in various (model-dependent) ways. First, one can compare the number of detected bursts to predicted rate of occurrence from a given model or to actual detections of expected counterparts (i.e. supernovae, in the case of long bursts, see e.g. \citealt{Soderberg2006}). Second, one can compare between early and late inferences of the energetics of the blast waves. At early times, assuming radial flow, the relevant energy is the energy per solid angle, or $E_{iso} / (4 \pi)$, where $iso$ stands for `isotropic equivalent'. If the jet subsequently spreads out sideways and becomes spherical (allowing eventually for the non-relativistic radial-flow self-similar limit), the energy per solid angle becomes $E_{jet} / (4 \pi)$, where $E_{jet} \approx E_{iso} \theta_0^2 / 2$, for small opening angles. Comparing between early and late time calorimetry should therefore yield opening angles (see e.g. \citealt{Berger2004, Shivvers2011}). Even if no late time calorimetry is possible, the often extremely high values for $E_{iso}$ resulting from early-time calorimetry (e.g. \citealt{Cenko2010}), already hint that the actual energies in the jets are likely lower. When collimation is accounted for (and jets with typical jet half opening angles of 6$^\circ$ are inferred, see e.g. \citealt{Racusin2009, Ryan2015}), the results tend to cluster around $E_{jet} \sim 10^{51}$ erg \citep{Bloom2003}.

The third way of inferring collimation, is by looking for signatures directly in the light curve. Two effects will lead to a steepening of the temporal evolution. On the one hand, the visible patch will grow as relativistic beaming weakens and reveal a lack of emission from beyond the edges once $\gamma \approx 1 / \theta_0$. This effect strongly depends on viewing angle, which puts the visible patch initially closer to the jet edge if moved off-axis \citep{vanEertenZhangMacFadyen2010, vanEertenMacFadyen2012}. On the other hand, the jet will start to spread out sideways. The over-pressured edges (relative to the circumburst environment) will do so immediately, and as the fact of the emergence of the jet into its environment is communicated towards the jet axis, more of the jet will follow. The broader jet will sweep up more material per unit time, leading to a faster deceleration. Because the spreading velocity of the jet is suppressed by a factor $1 / \gamma$ in the observer frame, this sets off a feedback loop where a slower jet is seen to sweep up even more material. Due to this $1/\gamma$ suppression factor from the Lorentz transform between the frame comoving with the blast wave and our frame, the sideways spreading of the jet becomes noticeable again once $\gamma \sim 1 / \theta_0$. If the jet were in full causal contact, the spreading behavior would be exponential \citep{Rhoads1999}.

No exact analytical solutions exist for spreading jets, even when starting from a `top-hat' conic wedge out of the spherical self-similar solution (once sphericity is dropped, initially structured and multi-component jets can be assumed too, see e.g. \citealt{Berger2003, Rossi2002}), although many toy models can be found in the literature (e.g. \citealt{Rhoads1999, SariPiranHalpern1999, KumarPanaitescu2000, Huang2000,  vanEertenZhangMacFadyen2010, Wygoda2011, GranotPiran2012}). Unfortunately, approximate models tend to be notoriously sensitive to the choices made for the underlying simplifications (see \citealt{Granot2007} for a discussion). The picture that has emerged from relativistic hydrodynamics (RHD) simulations (\citealt{ZhangMacFadyen2009, vanEertenZhangMacFadyen2010, Wygoda2011, DeColle2012,vanEertenMacFadyen2013}), is one where, for realistic opening angles (where $\theta_0 \ll 0.05$ rad does not apply), jet spreading does not achieve the runaway behavior and exponential increase in opening angle expected in the ultra-relativistic limit, but stays closer to logarithmic \citep{vanEertenMacFadyen2012}. A key reason for this is that full causal contact along all angles of the fluid takes time to establish, leaving little time in practice for an ultra-relativistic spreading regime because the blast wave quickly becomes trans-relativistic following the onset of spreading \citep{vanEerten2013}. As a result, the effect on the light curve following the `jet break', is due to the joint impact of both the edges becoming visible and the onset of spreading, with neither overwhelming the other. Post-break slopes remain dependent on observer angle and can be used as means to constrain jet orientation \citep{Ryan2015}.

\subsection{Reverse shocks and injection of energy into the flow}
\label{reverse_shock_section}

The preceding sections describe the subsequent evolution of an instantaneous explosion without initial mass content, i.e. a single forward shock moving into the circumburst medium. But unless the jet is driven nearly completely by pointing flux (see e.g. \citealt{lyutikov2006}), a certain amount of initial mass is expelled with the explosion. A simple way of incorporating this would be to add a mass $M_0$ to the shell model (but prior to deriving eq. \ref{shell_model_equation}, because the ejected mass is presumed to reside in a cold shell where all energy is converted into kinetic energy, while eq. \ref{shell_model_equation} only connects mass to shock-jump conditions). The presence of initial ejecta mass will postpone deceleration of the blast wave, which will initially coast along at fixed velocity in ballistic motion. A forward-reverse shock system is established, with the forward shock moving into the circumburst medium and the reverse shock heating up the ejecta. The width of the ejected shell influences the dynamics as well. Even an initially infinitesimally thin shell will stretch out to width $\Delta R \sim R / \Gamma^2$, similar to a decelerating blast wave. Only for shells wider than this, the initial shell width has to be taken into account explicitly when computing the deceleration radius, marking the turning point between coasting and decelerating of the blast wave. The two types of shell, wider or narrower than the intrinsic blast wave width, have been labeled `thick' and `thin' shells respectively in the literature \citep{SariPiran1995}. For thick shells, the reverse shock achieves relativistic velocity (in the ejecta frame) before crossing the ejecta. The pre-deceleration stage and reverse shock crossing were expected to be visible briefly (mainly in the optical), during the early evolution of the afterglow, for about $10^2$ s.

However, the Swift satellite \citep{Gehrels2004} has revealed instead the existence of an extended ($10^4$ s) early, flat phase of X-ray decay to be a common occurrence in afterglow light curves \citep{Nousek2006, ZhangBing2006}. A similar stage has been found to exist in the optical as well (see e.g. \citealt{PanaitescuVestrand2011, Li2012}), although a joint explanation for both X-rays and optical is complicated by the existence of very early steep decay in the X-rays (likely connected to the end of the prompt emission, rather than the afterglow), complex temporal optical emission profiles (maybe multi-component emission, from both reverse and forward shock) and the close proximity of the end of the shallow decay phase and of what is typically interpreted as a jet break \citep{Li2015}. If not explained from viewing angle effects (e.g. \citealt{EichlerGranot2006}) or evolving microphysics (e.g. \citealt{Granot2006, Filgas2011}), plateau stages point to some form of prolonged energy injection into the ejecta, where a continuous `push' from the back delays deceleration. This can take various forms, such as late catching up of slower material into the forward-shock / reverse shock system (e.g. \citealt{Panaitescu1998, ReesMeszaros1998, SariMeszaros2000}), long-term source luminosity (e.g. \citealt{ZhangMeszaros2001}) or conversion of Poynting flux from the ejecta (e.g. \citealt{Usov1992, Thompson1994}). In afterglow analysis, this injection can be modeled in the form of a power law increase in ejecta energy (e.g. \citealt{ZhangBing2006, Racusin2009}). In the case of a relativistic reverse shock (in the frame of the inflowing material) and gradual, power-law type injection, one can even maintain self-similarity \citep{Blandford1976, vanEerten2014injection}. Another promising modeling approach which has been applied to data directly, is dropping self-similarity, but maintaining a simplified description for the late shells \citep{Uhm2012, DePasquale2015}. It should be emphasized that, although long-term engine activity is certainly a possible explanation for these early stages (potentially requiring a magnetar engine model, see e.g. \citealt{Usov1992, Thompson1994, DaiLu1998, ZhangMeszaros2001}), jet breakout is a messy event (see e.g. \citealt{WaxmanMeszaros2003, MorsonyLazzatiBegelman2007}) which might well naturally introduce an extended observable early stage even for briefly active engines, before moving towards the asymptotic regime of a decelerating relativistic blast wave \citep{DuffellMacFadyen2014}. 

A potential means of distinguishing between engine models are the correlations found between plateau end times and luminosity in the X-rays \citep{Dainotti2008}, and in the optical \citep{PanaitescuVestrand2011, Li2012}. The two correlations have different slope, which is consistent with the optical and X-ray emission typically being observed to be in different spectral regimes \citep{Greiner2011}. They emerge naturally from a synchrotron forward-reverse shock system \citep{Leventis2014, vanEerten2014injection}, but require the presence of a relativistic reverse shock \citep{vanEerten2014}, or, `thick' shells rather than `thin' shells. It is not clear how strong the emission from the reverse shock will be in reality, since the relative strength of reverse shock emission is sensitive to model assumptions, such as the degree of magnetization of the ejecta (see e.g. \citealt{MimicaGianniosAloy2009}), and can vary wildly even for a standard synchrotron model (e.g. \citealt{Leventis2014}).

\subsection{Further complications}

As already alluded to above, the initial geometry of the ejecta and the environment of the burster provide two obvious complications to the standard picture. Even the direct environment of the burster can reasonably be expected to have a complex shape. The stellar wind profile will only extend to a finite range and be influenced by photo-ionization, stellar rotation and fluid instabilities \citep{Eldridge2006, vanMarle2006, Eldridge2007, vanMarle2008, vanMarleKeppens2012}. Late time mass loss of the progenitor system is likely erratic \citep{Mesler2012}. Although circumburst mass transitions are not expected to introduce sudden changes in the observed light curves from the forward shock \citep{NakarGranot2007, vanEerten2009, GatvanEertenMacFadyen2013, Geng2014}, an overall slope transition can reasonably be expected, which could explain $k$ measurements other than $k=0$ or $k=2$ \citep{Curran2009}. Additional emission might be generated by a complex shock structure following multiple interactions \citep{UhmZhang2014, Mesler2014}.

\section{Emission}
\label{emission_section}

\begin{figure}
 \centering
  \includegraphics[width=0.95\columnwidth]{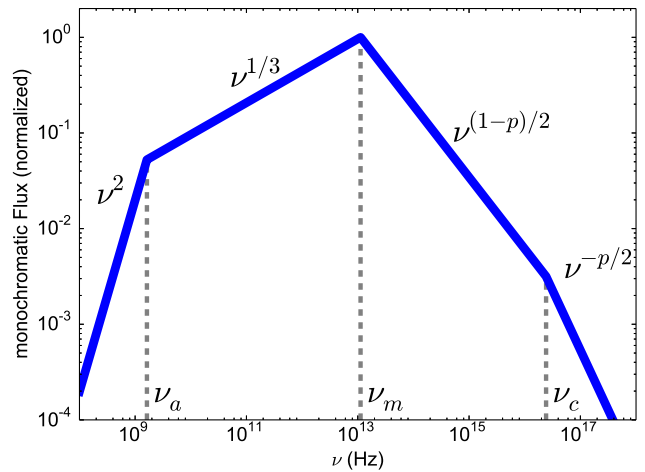} \\
\caption{Typical synchrotron spectrum in the slow cooling case.}
 \label{spectrum_amc_figure}
\end{figure}

The synchrotron spectrum consists of a series of connected power laws, separated by break frequencies, and evolves in a characteristic manner during the lifetime of the blast wave \citep{MeszarosRees1997, Wijers1997, SariPiranNarayan1998}. An example is provided by Fig. \ref{spectrum_amc_figure}. We will discuss this \emph{slow cooling} case first. A shock-accelerated electron population $n_e(\gamma_e)$ with power-law $p \sim 2.3$ is typically assumed, with $n_e(\gamma_e) \propto \gamma_e^{-p}$ and $\gamma_e$ expressed in the comoving fluid frame. The lower cut-off value for this distribution is $\gamma_m$. If we further parametrize the shock-accelerated electron spectrum using $\xi_N$ (typically taken $\sim 1$), the fraction of electrons that get shock-accelerated, and $\epsilon_e$ (typically $\sim 0.1$), the fraction of available internal energy in the fluid that goes into the non-thermal electron population, one can derive (by equating integrals over electron number density distribution and electron energy density distribution $\gamma_e n_e(\gamma_e) m_e c^2$ to available total number density and energy density respectively):
\begin{equation}
\gamma_m = \frac{2 - p}{1 - p} \left( \frac{\epsilon_e}{\xi_N} \frac{e}{\rho} \frac{m_p}{m_e c^2} \right),
\end{equation}
where $m_p$ proton mass and $m_e$ electron mass. If the power law distribution slope $p$ is too shallow ($p < 2$), one can choose to either maintain a physically plausible proportionality between $\gamma_m$ and $e / n$, or have the upper cut-off to the particle population (which can be ignored for $p>2$) dictate $\gamma_m$ instead, in order to maintain the interpretation of $\epsilon_e$ \citep{Bhattacharya2001, DaiCheng2001}.

According to synchrotron theory, the local spectrum from an electron at energy $\gamma_e$ peaks around
\begin{equation}
\nu'_{e} \approx \frac{3}{4 \pi} \gamma_e^2 \frac{q_e B}{m_e c},
\end{equation}
in the frame comoving with the fluid\footnote{This choice of notation was made for consistency with the literature. Note that we now have $\nu'$ and $\rho$ in the frame comoving with the fluid, while $\nu$ and $\rho'$ are expressed in the lab frame.} and where $q_e$ electron charge. Magnetic field $B$ is typically parametrized again via a fraction of available energy, according to $B^2 / (8 \pi) \equiv \epsilon_B e$ (and with $\epsilon_B$ typically $\sim 0.01$). The critical frequency $\nu_m$ shown in Fig. \ref{spectrum_amc_figure}, represents the \emph{average} critical frequency for the combined emission of all local synchrotron spectra and their local $\gamma_m$ values, and expressed in the observer frame. The spectral slopes of $1/3$ and $(1-p)/2$ at both sides of $\nu_m$ also follow from standard synchrotron theory (see e.g. \citealt{RybickiLightman1979}).

The dependency of the flux on the model parameters $E_{iso}$, $\rho_{ext}$, $\epsilon_B$, $\epsilon_e$, $\xi_N$, $z$, $d_L$ (luminosity distance) in a given spectral regime can now be determined as follows. The emission coefficient peaks according to synchrotron theory at 
\begin{equation}
\epsilon'_{base} \sim \frac{p-1}{2} \frac{\sqrt{3} q_e^3}{m_e c^2} \xi_N n B,
\end{equation}
in the frame comoving with the fluid and where $n$ the local comoving fluid number density (such that $\xi_N n$ the number density of non-thermal electrons). On both sides of $\nu'_{m}$, we have
\begin{eqnarray}
\epsilon'_{\nu} & = & \epsilon'_{base} \left( \frac{\nu'}{\nu'_{m}} \right)^{1/3}, \quad \nu' < \nu'_{m} < \nu'_{c}, \nonumber \\
\epsilon'_{\nu} & = & \epsilon'_{base} \left( \frac{\nu'}{\nu'_{m}} \right)^{(1-p)/2}, \quad \nu'_{m} < \nu' < \nu'_{c}, 
\label{emission_m_equation}
\end{eqnarray}
where $\nu'$ the observer frequency and $\nu'_{c}$ the cooling break in the frame comoving with the fluid, which we will discuss below. In our frame $\epsilon_{\nu'} \approx \gamma^2 \epsilon'_{\nu'}$, as the dependency evolution of flux on model parameters will be dictated by emission from material moving (nearly) straight towards the observer. The observed flux is then
\begin{equation}
F_{\nu', \oplus} \propto \frac{1+z}{d_L^2} V \epsilon_{\nu'}.
\label{F_thin_equation}
\end{equation}
Here $V$ is the emitting volume, the product of area $(R/\gamma)^2$ and depth $\Delta R \propto R / \gamma^2$. The Lorentz factor in the area reflects the size of the visible patch due to beaming. Without sideways spreading, it would be sufficient to omit this factor in order to obtain post jet-break flux values. Spreading models quickly become more complicated (simulation-based post-break light curve slopes for the ISM case are provided by \citealt{vanEertenMacFadyen2013}. These simulations reveal steep temporal slopes of $\sim-2.7$ above $\nu_m$, once the transition to the post-break regime has completed, which is strikingly steeper than indicated by the Swift sample). Observed frequencies are related to comoving frequencies via the usual $\nu_{\oplus} \approx \gamma \nu' / (1+z)$, and observed time and emission time via eq. \ref{obs_time_equation}. Using the jump-condition values from eq. \ref{shock_jump_equation} and the dynamics from eq. \ref{shell_model_equation}, is then sufficient to determine the exact dependence of flux on the model parameters. Synchrotron flux equations for afterglow blast waves can be found at various places in the literature, including extensions such as trans-relativistic flow, energy injection and general values of $k$ (see e.g. \citealt{MeszarosRees1997, SariPiranNarayan1998, WaxmanKulkarniFrail1998, GruzinovWaxman1999, GranotSari2002, vanEertenWijers2009, Leventis2012, YiWuDai2013, Gao2013, vanEerten2014injection}).

If not all available electrons are shock-accelerated into a non-thermal population, i.e. $\xi_N < 1$, some will remain in a Maxwellian distribution. This can potentially impact the synchrotron light curve. Unfortunately, $\xi_N$ can not be derived from observations of the power-law electrons directly, even if all spectral regimes could be observed. A full degeneracy between $\xi_N$ and other model parameters exist, where a decrease in $\xi_N$ can be compensated for by a simultaneous linearly proportional decrease in $\epsilon_e$, $\epsilon_B$ and linearly proportional increase in $E_{iso}$ and $\rho_{ext}$ \citep{EichlerWaxman2005}.

\subsection{Electron Cooling}

\begin{figure}
 \centering
  \includegraphics[width=0.95\columnwidth]{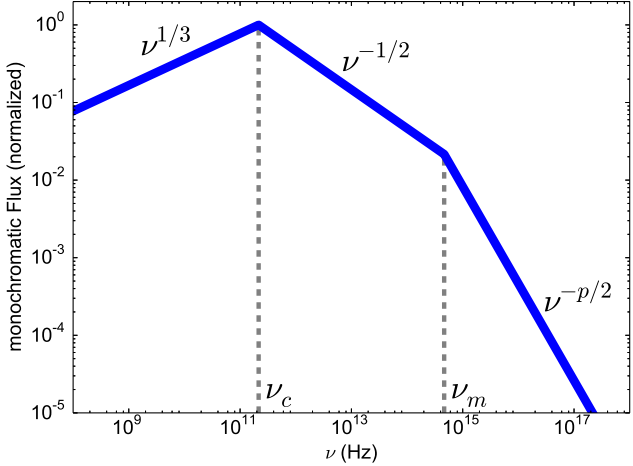}
\caption{A fast cooling synchrotron spectrum.}
 \label{spectrum_cm_figure}
\end{figure}

Another characteristic of the synchrotron spectrum is provided by electron cooling, since energetic particles use their energy very quickly through synchrotron emission:
\begin{equation}
 \frac{\dev \gamma_e}{\dev t'} = - \frac{4 \sigma_T \gamma_e^2}{3 m_e c} \epsilon_B e + \frac{\gamma_e}{3 n} \frac{\dev n}{\dev t'},
\label{kinetic_equation}
\end{equation}
where $\sigma_T$ the Thomson cross-section and $t'$ in the comoving fluid frame. Beyond the cooling break $\nu_c$, the effect of cooling becomes important and the first term on the right above (the synchrotron loss term) dominates. Below the cooling break, the other term (cooling through adiabatic expansion, note the `stretching' $\dot{n}/n$) dominates. The cooling break $\nu_c$ can lie either above or below $\nu_m$, yielding spectra designated as slow- and fast-cooling respectively.

A simple approaching to modeling the behavior of the cooling break, is to ignore the spatial structure of the blast wave for the purpose of determining the cooling time, using a steady-state approximation to eq. \ref{kinetic_equation} and equating overall cooling time to burst duration, resulting in 
\begin{equation}
\gamma_c = \frac{6 \pi m_e c \gamma}{\sigma_T B^2 t},
\end{equation}
connected to a characteristic frequency in the usual manner via $\nu_c \propto \gamma_c^2 B$ \citep{SariPiranNarayan1998}. This approach has been used in simulations as well (e.g. \citealt{ZhangMacFadyen2009, DeColle2012, vanEertenvanderHorstMacFadyen2012}), even though other quantities ($\nu_m$, peak flux) are calculated completely locally. The correct scalings and temporal evolution are reproduced in this manner, but this hybrid approach (in the simulation case), implies an offset of the cooling break relative to a fully local approach to cooling. The reason for this is that the full fluid profile provides a dimensionless constant of integration when computing the cooling break from local spectra (which exhibit an exponential cut-off locally, rather than a power-law transition, only globally adding up to such a transition) that is different from the one provided by an effectively flat fluid profile. The good news is that this off-set remains essentially constant throughout the evolution of the blast wave \citep{vanEertenZhangMacFadyen2010}, but the effect on multi-band analysis of afterglows can be substantial \citep{Guidorzi2014}. A local approach to electron cooling in simulations (e.g. by rewriting eq. \ref{kinetic_equation} in the form of an advection equation for $\gamma_m$), requires extreme resolutions, which is challenging already in one dimension \citep{vanEerten2010transrelativistic}, but only achievable in multi dimensions by specialized methods (e.g. \citealt{vanEertenMacFadyen2013}).

In the global approach, the emission coefficient equations \ref{emission_m_equation} can be extended to include the effect of electron cooling as indicated in Figs. \ref{spectrum_amc_figure} and \ref{spectrum_cm_figure}, leading to
\begin{eqnarray}
\epsilon'_{\nu} & = & \epsilon'_{base} \left( \frac{\nu'_c}{\nu'_{m}} \right)^{(1-p)/2} \left( \frac{\nu'}{\nu'_{c}} \right)^{-p/2}, \quad \nu'_m < \nu'_{c} < \nu', \nonumber \\
\epsilon'_{\nu} & = & \epsilon'_{base} \left( \frac{\nu'}{\nu'_{c}} \right)^{-1/2}, \quad \nu'_{c} < \nu' < \nu'_{m}. 
\end{eqnarray}
Here $\epsilon'_{base}$ tracks the peak of the spectrum, which not necessarily coincides with $\nu_m$. In the local approach to cooling, this only describes the shape of the globally emergent spectrum (i.e. flux, rather than emission coefficients). The $-1/2$ power can be understood as follows. The frequencies in this regime probe electron Lorentz factors below the injection value of $\gamma_m$, while the cooling timescales are extremely short since we are above $\nu_c$ (meaning that the shape of the injected profile is not relevant, so no $p$-dependency). All available energy is quickly radiated away. Per frequency, this yields $F_\nu \propto \gamma_e m_e c^2 / [\gamma_e^2 B] \propto \gamma_e^{-1} \propto \nu^{-1/2}$.

\subsection{Synchrotron Self-absorption}

\begin{figure}
 \centering
  \includegraphics[width=0.95\columnwidth]{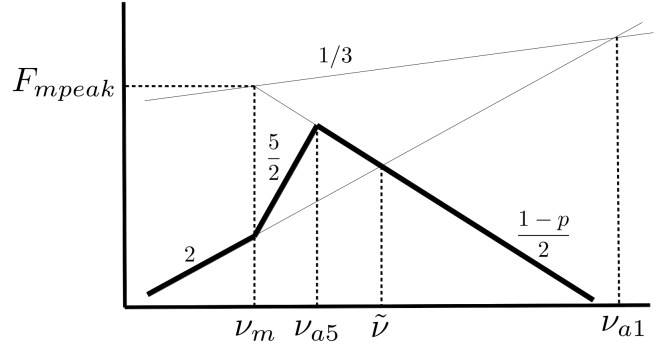} \\
\caption{self-absorbed spectrum beyond $\nu_m$, in terms of regular ordering for $\nu_a < \nu_m$.}
 \label{spectrum_am_figure}
\end{figure}

At low frequencies (typically in the radio regime), synchrotron self-absorption (ssa) becomes important and a characteristic break $\nu_a$ appears in the spectrum, below which the flux drops off more steeply. Since ssa is not a scattering process, it is relatively straightforward to model in a linear radiative transfer approach, \citep{GranotPiranSari1999, GranotSari2002} that can also be applied to adiabatic relativistic blast wave simulations \citep{Mimica2009, vanEerten2010transrelativistic}. In the absence of electron cooling, the absorption coefficient due to ssa is given by
\begin{equation}
\alpha'_{\nu'} \sim (p-1) (p+2) \xi_N n \frac{ \sqrt{3} q_e^3 B}{\gamma_m 16 \pi m_e^2
c^2} (\nu')^{-2} \left( \frac{ \nu'}{\nu'_m} \right)^{\kappa},
\label{alpha_equation}
\end{equation}
where $\kappa = 1/3$ if $\nu' < \nu'_m$ and $\kappa = -p/2$ otherwise. In a simplified computation in order to obtain the model parameter dependencies, we need to consider an emitting surface $A \propto (R/\gamma)^2$ and a source function below $\nu_a$, rather than the emitting volume for the optically thin case \citep{SariPiranNarayan1998, WaxmanKulkarniFrail1998}, i.e.
\begin{equation}
F_{\nu', \oplus} \propto \frac{1+z}{d_L^2} A \frac{\epsilon_{\nu'}}{\alpha_{\nu'}}.
\label{F_a_equation}
\end{equation}
In the lab frame $\alpha_{\nu'} \approx \alpha'_{\nu} / \gamma$. In the slow cooling case and with $\nu_m > \nu_a$, the spectral slope $2$ below $\nu_a$ (see Fig. \ref{spectrum_amc_figure}) follows from a comparison between eqs. \ref{alpha_equation} and \ref{emission_m_equation}. In case $\nu_m$ and $\nu_a$ flip, a new slope of $5/2$ is introduced, as can be seen from the same equations. Once $\nu_a > \nu_m$, the temporal evolution of $\nu_a$ will change, as will the peak of the spectrum, which now occurs at $\nu_a$, rather than $\nu_m$. Nevertheless, these differing characteristics follow from the regular self-absorption and $\epsilon'_{base}$ evolution, as can be seen geometrically in Fig. \ref{spectrum_am_figure}, where $\nu_{a,1}$ the position when extrapolating the $\nu_m > \nu_a$ case and $\nu_{a,5}$ the actual position of the self-absorption break (see also \citealt{Leventis2012} for a discussion. The numbers 1 and 5 were chosen to match the notation from \citealt{GranotSari2002}). The model parameter dependency and evolution of $\nu_a$ (i.e. $\nu_{a,1}$) can be determined by looking for the meeting point between the flux according to eq. \ref{F_a_equation} and eq. \ref{F_thin_equation}, using $\nu < \nu_m$ for both.

As said, the case $\nu_{a,5} > \nu_m$ does not introduce anything new, although the flux equations look different. This minor observation has the practical implication that, when constructing scale-invariant spectral templates from high-resolution simulations, all that is needed are the temporal evolution curves for $\nu_m$, $\nu_c$, $\nu_{a,1}$ and $F_{peak}$, with the understanding that the latter does not coincide with the actual spectral peak once $\nu_{a,5} > \nu_m$.

The fact that self-absorption renders only the front of the blast wave visible to the observer should serve as caution when attempting to seek out early time emission from a reverse shock in the radio domain: depending on density and profile of the environment, this component might well be hidden from view by ssa. A precise analysis of the early stage radio emission is further complicated by the role of electron cooling. At early time, we might also be observing the fast cooling case, rather than the slow cooling case, and the cooling break will help shape the absorption coefficient. At this point the difference between local and global cooling emerges again as well. An exact treatment of local cooling will actually introduce additional spectral regimes that are not apparent in a global approximation. These topics are discussed further by \cite{GranotPiranSari1999, GranotSari2002}.

\subsection{The sharpness of spectral breaks}

The connected power law description of the synchrotron spectrum that we applied above, is of course an approximation. In reality, the asymptotic regimes approach one another smoothly, with the underlying shape for a single electron spectrum being an integrated modified Bessel function of fractional order. Since the full expression for a spectral transition is cumbersome and, more importantly, since the spectral sharpness also depends on the fluid configuration that shapes all the local contributions to the emergent spectra, it is more convenient in practice to use approximate formulae, typically smooth power laws of the type
\begin{equation}
Y(x) = Y_0 \left[ \left( \frac{x}{x_0} \right)^{-s \alpha_1} + \left( \frac{x}{x_0} \right)^{-s \alpha_2} \right]^{-1/s},
\end{equation}
and varieties thereof, with the most useful shape depending on e.g. the sign of the transition $\alpha_2 - \alpha_1$. The larger $s$, the sharper the transition. Such approximations can be applied to temporal transitions (e.g. jet breaks) as well and have been applied to both light curves and spectra (e.g. \citealt{Beuermann1999, Harrison1999, GranotSari2002, vanEertenWijers2009, Leventis2012, vanEertenMacFadyen2013}). Smooth power laws can lead to significantly better fits to data, and the flux exactly at a spectral transition can differ up to an order of magnitude from sharply connected power laws. The exact value of $s$ is hard to determine in practice from the data, and good fits can typically be obtained for a range of values. The theoretical values of $s$ are influenced by many things. In addition to the fluid profile, the closeness of other spectral transitions also plays a role even if their presence is not immediately apparent from the data. Finally, the sharpness of the spectrum is sensitive to the orientation of the jet, since a given observer time corresponds to a different set of emission times for each angle \citep{vanEertenMacFadyen2012scaling}.

\subsection{Further complications}

Aside from the complexities introduced by more realistic particle spectra and by multiple emission sites for synchrotron emission (e.g. a forward shock and a reverse shock), the most obvious further complication is the addition of other radiative processes. Of these, inverse Compton scattering and synchrotron self-Compton scattering are the most likely candidates. Inverse Compton scattering has a noticeable impact on the cooling of electrons when $\epsilon_B \ll \epsilon_e$ \citep{SariNarayanPiran1996, PanaitescuKumar2000, SariEsin2001}. This effect can be included in prescriptions for synchrotron spectra. Even when not observed directly, Inverse Compton scattering will shift the cooling break downwards, and, in the fast cooling stage, the self-absorption break upwards. \citep{SariEsin2001, GranotSari2002}.

Other factors that complicate interpreting the observed emission in terms of a synchrotron blast wave origin, is the contribution from completely separate components, such as host galaxies and supernovae, or dust echoes of the prompt emission \citep{Evans2014}.

\section{Model-based data analysis}
\label{fitting_section}

When analyzing the data, a number of approaches can be taken. The conventional approach has been to start from analysis of light curves and spectra (when available) in terms of simplified heuristic fit functions, typically power laws, augmented where appropriate by descriptions of extinction and absorption due to intervening material and host galaxy and supernova flux contributions. This approach results in a concise description of the data (in itself already useful) that can subsequently be interpreted in terms of physical models, under the assumption that the essence of these models can be captured sufficiently in the form of power laws. On the other hand, once can test physical models against the data directly in a manner that does not require synchronous multi-band observations (e.g. \citealt{PanaitescuKumar2001, PanaitescuKumar2002, Yost2003}), which has become increasingly popular in recent years (e.g. \citealt{vanEertenvanderHorstMacFadyen2012, Leventis2013, Laskar2013, Perley2014, Urata2014, Guidorzi2014, Ryan2015, ZhangBinbin2015}). Rather than yielding the `best' short-hand description of the data set, this immediately addresses the question of whether a given physical model can explain it, and to the extent that it can, provides estimates for the model parameters. An additional advantage of this approach is that arbitrarily complex light curve shapes (as provided by the underlying model) can be accounted for.

These model fitting approaches also naturally connect to probabilistic data analysis methods, including Bayesian inference, which are having a transformative effect on the field. Software packages that implement numerical methods such as affine invariant Markov-Chain Monte Carlo (MCMC, \citealt{GoodmanWeare2010, Foreman-Mackey2013}) or multi-modal nested sampling \citep{Feroz2009, Buchner2014} are becoming readily available and are ideally suited to GRB afterglow analysis, due to their capability of dealing efficiently with expensive fitting functions, large numbers of free parameters and bimodal posterior distributions. The possibility in Bayesian data analysis of marginalizing over nuisance parameters is extremely valuable for assessing the performance of a model in case not all parameters can be constrained fully (as is often the case for afterglows, since this would require broadband data covering all spectral regimes of the synchrotron spectrum; such data sets exist, but are still rare, although the sample is growing). 

\subsection{Simulations}

\begin{figure}
 \centering
  \includegraphics[width=0.95\columnwidth]{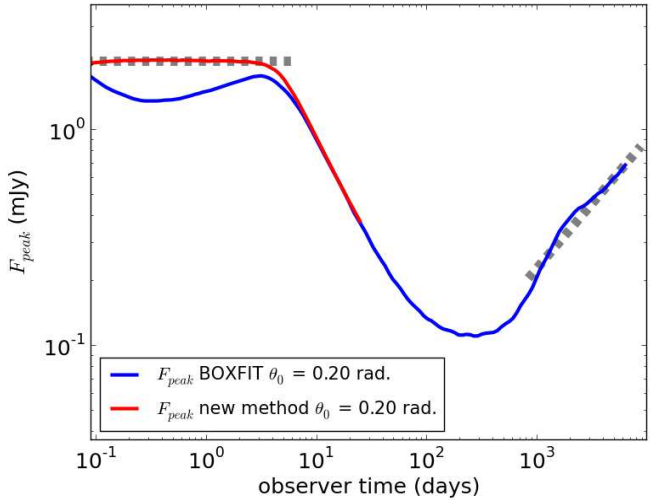} \\
\caption{Evolution of the synchrotron spectral peak, computed from lab-frame (blue) and boosted-frame (red) simulations. Dashed grey lines denote the asymptotic self-similar limits. The drop between $10^0$ - $10^1$ days describes the effect of a jet-break. The additional bump at late time the rise of the counterjet.}
 \label{simulation_res_figure}
\end{figure}

Because the deceleration to trans-relativistic velocities and the spreading behavior of relativistic blast waves are difficult to capture analytically, many groups have used RHD simulations in one and more dimensions to study this crucial stage of afterglow blast wave evolution (including, but not limited to \citealt{Kobayashi1999, Downes2002, CannizzoGehrelsVishniac2004, Meliani2007, ZhangMacFadyen2009, vanEerten2011chromaticbreak, Wygoda2011, DeColle2012Mezcal, vanEertenMacFadyen2013, DuffellMacFadyen2013}). Most multi-dimensional simulations employ special strategies to deal with the stringent resolution requirements (i.e. $\Delta R \sim R / \Gamma^2$, with $\Gamma \gg 1$), specifically adaptive-mesh refinement, where the grid resolution is locally dynamically doubled (or halved) along each dimension, depending on the variability of the flow. With this approach, the six orders of magnitude between initial shell width and final shell radius could be more or less covered numerically. However, it is important to note that \emph{all} simulations prior to 2013 failed to fully resolve numerically the blast wave at the key dynamical stage of jet spreading. That qualitatively correct behavior was nevertheless reproduced, could only be confirmed recently \citep{vanEertenMacFadyen2013} by simulations in a specialized Lorentz-boosted frame (a natural antidote against $1/\Gamma^2$ related issues). These resolution issues also impact light curves computed from simulations. The strategy for such computations is not dissimilar from inferring flux from simplified models: take the dynamics as a given, assuming adiabatic expansion, and employ radiative transfer or, above the self-absorption break, perform a straightforward summation of locally emitted power. As described previously, the time-dependent synchrotron spectrum can be captured concisely based on its key characteristics. An example is provided by Fig. \ref{simulation_res_figure}, showing both lab-frame and boosted frame evolution curves for the spectral peak.

Because derived synthetic light curves contain emission from many different emission times arriving at each single observer time, the resolution issue is in practice not problematic at the light curve level for homogeneous medium simulations: the resolution issue is at its most severe early on during the simulation, and the emission from these times is observed jointly with emission preceding the simulation starting time (i.e. from the analytical self-similar Blandford-McKee solution that provides the initial conditions for the simulation). Only when jets become too narrow ($\theta_0 \ll 0.05$ rad), observer frequencies drop too far below $\nu_{a}$ at early times, or for certain circumburst medium profiles (including, unfortunately, stellar wind), this becomes a potential issue. 

\subsection{Scale-invariance and model templates}

The complete evolution of a blast wave, from early time conic wedge out of the relativistic self-similar solution to the spherical non-relativistic self-similar solution, exhibits a useful scale invariance that can be employed for model comparisons to data \citep{vanEertenvanderHorstMacFadyen2012}. This invariance goes beyond self-similarity and follows straightforwardly from dimensional analysis. The fluid equations can be expressed in terms of spacetime coordinates $\mathcal{A} \equiv rc / t$, $\xi$, $\theta$, rather than $r$, $t$, $\theta$. In the non-relativistic limit $\mathcal{A} \downarrow 0$, while $\theta$ no longer applies due to sphericity, and the self-similar solution of single variable $\xi$ is recovered. In the ultra-relativistic limit $\mathcal{A} \uparrow 1$, while $\theta$ again does not apply due to the radial flow assumption, leaving again a self-similar solution. Note that $\chi = \chi(\xi)$ and the Blandford-McKee self-similarity variable is more practical in this limit. But even for intermediate values of $\mathcal{A}$, it remains true that rescaling in explosion energy or circumburst density can be compensated for with a rescaling of the coordinates. Taking $E_{iso} \to \kappa E_{iso}$, $\rho_{ref} R_{ref}^k \to \lambda (\rho_{ref} R_{ref}^k)$, $r \to (\kappa / \lambda)^{1/(3-k)} r$ and $t \to (\kappa / \lambda)^{1/(3-k)}$, leads to the invariance $\mathcal{A} \to \mathcal{A}$, $\xi \to \xi$ and $\theta \to \theta$: a bigger explosion explosion (or one in a more dilute medium) goes through the exact same stages as a smaller one, albeit at larger radii and at later times. In terms of dimensions, we scaled grams by a factor of $\kappa$, and centimeters and seconds both by a factor $( \kappa / \lambda )^{-3/(3-k)}$. The implied scalings for mass densities, energy densities and pressure are identical: $\rho \to \kappa^{-k / (3-k)} \lambda^{3 / (3-k)} \rho$, $e \to \kappa^{-k / (3-k)} \lambda^{3 / (3-k)} e$, $p \to \kappa^{-k / (3-k)} \lambda^{3 / (3-k)} p$. Lorentz factors remain unaffected, $\gamma \to \gamma$.

The big practical relevance of this comes when building a template set out of simulations, for comparison to observational data. Two dimensions in parameter space are now accounted for ($E_{iso}$ and $\rho_{ext}$, leaving only $\theta_0$). And, although we have no exact analytical solution for the spreading stage, the blast wave nevertheless segues smoothly from one asymptote towards the other. This means that we can use our intuition from the self-similar asymptotes (with features such as e.g. $\Delta R \propto R / \Gamma^2$) to guide us towards a suitable compression algorithm for simulation data, rendering the construction of templates feasible even if the original simulations are very computationally intensive. In this manner, model results along the remaining $\theta_0$ dimension in parameter space can be tabulated concisely \citep{vanEertenvanderHorstMacFadyen2012}. It should be noted that introducing energy injection does not break dynamical scale invariance, even though it introduces extra dimensions in parameter space, such as energy injection duration, that scale along \citep{vanEerten2014}.

Although dynamical templates are in principle sufficient for simulation-based afterglow analysis (with the caveat that radiative transfer based on dynamical templates needs to be calculated on-the-fly), we can do better if the radiative process of interest is sufficiently simple: a convenient feature of the power law nature of the synchrotron spectrum is that it renders it scale invariant \citep{vanEertenMacFadyen2012scaling}. Like the dynamical scale-invariance, this follows directly from dimensional analysis, and although in synchrotron spectra a number of additional constants appear (such as $m_e$), this invariance is not compromised within a single power law regime. In the flux equation, the role of fractions $\epsilon_B$, $\epsilon_e$ and $\xi_N$ does not change over time either. All this implies that evolution curves for $\nu_m$ and other characteristics  of the synchrotron spectrum (peak flux, cooling break, self-absorption break), even when derived from high-resolution multi-dimensional simulations, can be rescaled between model parameters, allowing for a synchrotron spectral template-based approach to afterglow fitting.

\subsection{\textsc{boxfit} and \textsc{scalefit}}

\begin{figure}
 \centering
  \includegraphics[width=0.99\columnwidth]{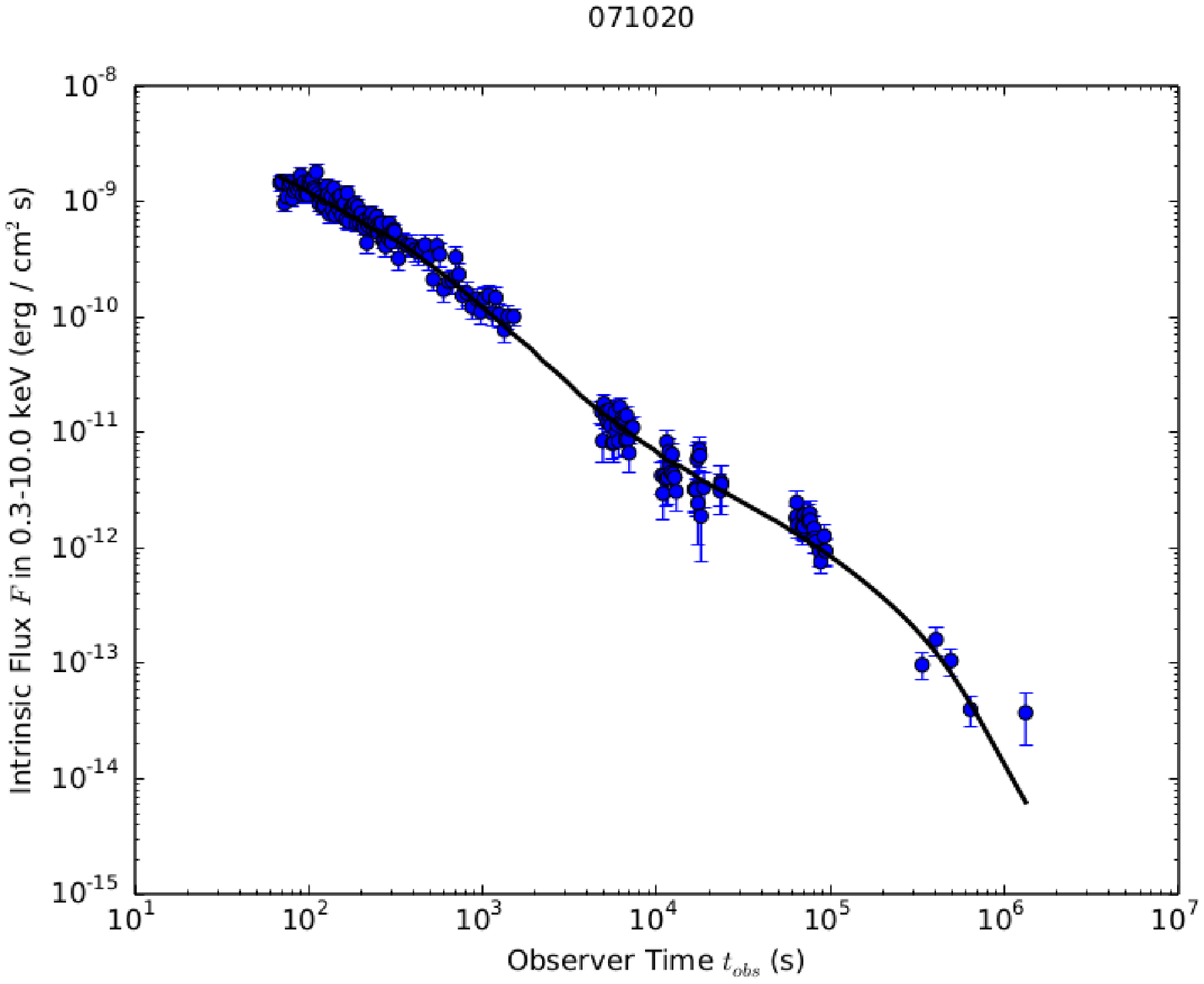} \\
  \includegraphics[width=0.90\columnwidth]{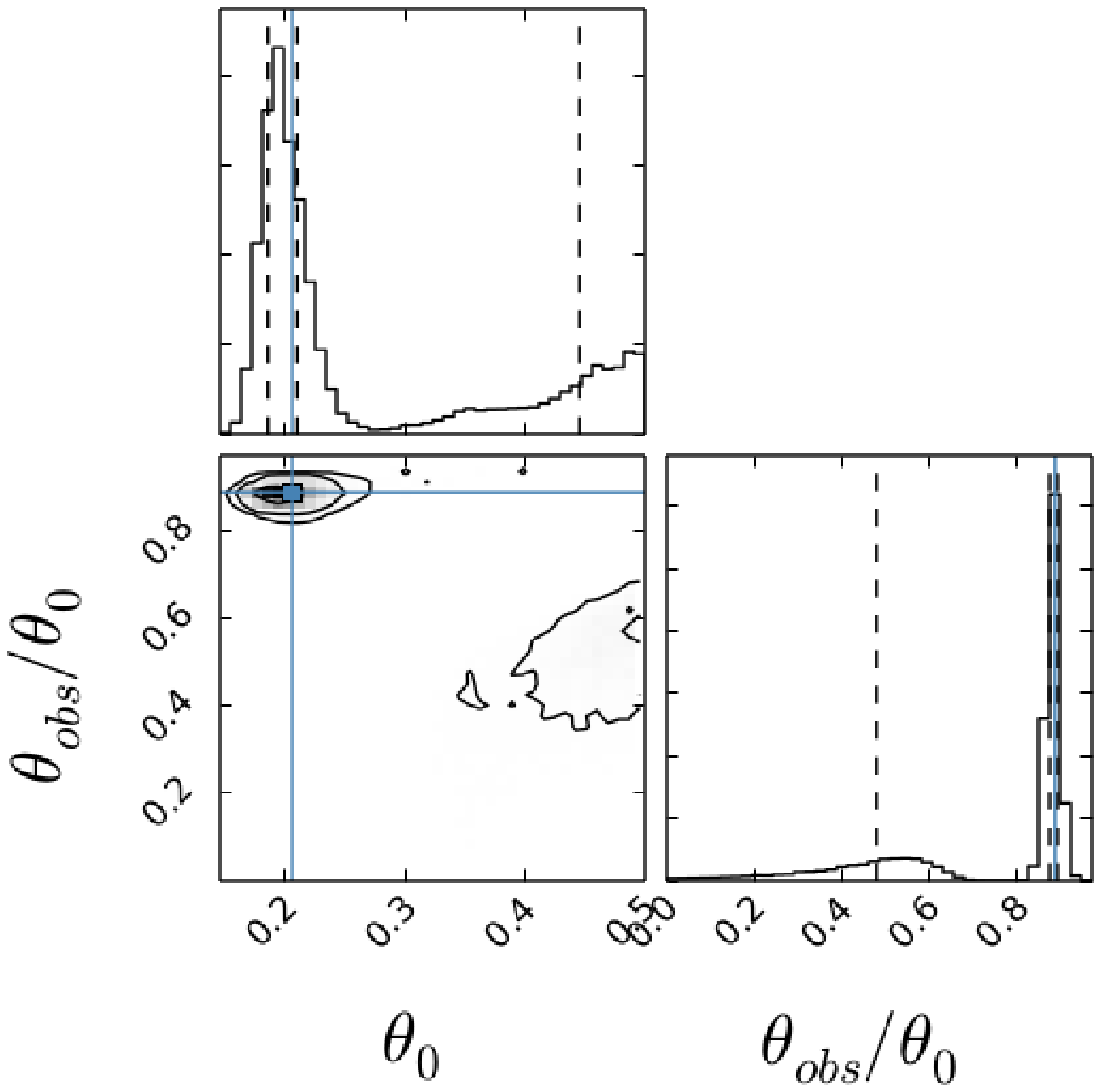} \\
\caption{A spectral template-based fit to the X-ray afterglow for GRB 071020, from \cite{Ryan2015}. Top figure shows the best fit, bottom figure the posterior probability distribution projections for jet opening angle and observer angle, marginalized over the other fit parameters ($E_{iso}$, $\rho_{ext}$, $p$, $\epsilon_B$, $\epsilon_e$). In a single band fit, most parameters remain unconstrained.}
 \label{fit_result_figure}
\end{figure}

The scale-invariances described above have been used to prepare synchrotron templates directly from relativistic hydrodynamics simulations \citep{vanEertenvanderHorstMacFadyen2012, vanEertenMacFadyen2012scaling, Ryan2015}. A simulation-based analysis code utilizing a set of 19 homogeneous medium simulations and dynamical scale invariance (\textsc{boxfit}, \citealt{vanEertenvanderHorstMacFadyen2012}), is freely available for download\footnote{\textsc{http://cosmo.nyu.edu/afterglowlibrary}}. Its follow-up, using scale-invariance at the level of spectral templates directly, is currently in preparation (\textsc{scalefit}; a first application to Swift XRT data can be found in \citealt{Ryan2015}, limited to homogeneous medium simulations) and will be available in the near future via the same website as \textsc{boxfit}. The full release of \textsc{scalefit} will include spectral templates based on approximately $70+$ high-resolution simulations including `boosted Lorentz frame' based simulations \citep{vanEertenMacFadyen2013} and a stellar-wind environment.

When emission is computed from radiative transfer through a fluid dynamical solution provided either analytically, from simulations directly, or from templates, the resulting spectra will be smooth automatically. Direct reconstruction of spectra from scale-invariant templates, on the other hand, requires that this smoothness is either ignored or accounted for explicitly: spectral sharpness itself is not scale-invariant, because it connects two regimes where the flux equations scale \emph{differently} for changes in $E_{iso}$ and $\rho_{ext}$. Additionally, there is the practical issue of how to deal with crossings of spectral breaks \citep{Leventis2012}. 

Another complication is that the physical models describe spectral flux, while detectors such as Swift/XRT count photons. A transition from count space to flux space needs to be made \citep{Evans2009}, while at the same time absorption (in X-rays) and extinction (in the optical), need to be accounted for. This transition is not completely trivial, and for example requires that some choice be made for the underlying spectral shape in the relevant spectral range. Ideally, the spectral shape is provided by the physical model as well, and the fit is essentially done in count space. The disadvantage of this approach is that this renders the fit software instrument-specific.

\section{Conclusions}

Gamma-ray burst afterglows provide a unique opportunity to study highly relativistic flows and fundamental plasma kinetic theory in a setting that is impossible to repeat on earth. Their core ingredients, relativistic blast waves and non-thermal emission from shock-accelerated particles, can be well understood from simplified models, a few of which have been described in detail in this paper. Nevertheless, GRB afterglows exhibit a rich range of features that follow both from the interplay of the standard model components and from the inevitable complications that can be added to the basic models. Jet and blast wave dynamics present an example of this, and it has only recently become possible to accurately model the evolution of GRB outflows, through high-resolution numerical hydrodynamics simulations, in a manner that can be compared directly to broadband observations. Even then, `accurate' has to be understood in terms of the simplicity of the initial assumptions for the jets, and more complicated initial jet profiles will (at least initially), give rise to more complicated light curves.

The two codes mentioned by name, \textsc{boxfit} and \textsc{scalefit}, offer the means to perform this type of simulation-based model fitting, but much work remains to be done. An extensive set of templates is being generated including stellar-wind type circumburst profile, and the public release for \textsc{scalefit} is in preparation. The strengths of these model-based codes are that they allow for direct model testing and fitting of complicated light curve shapes, incorporate advanced statistical methods and avoid the need for simultaneous broadband observations. The obvious disadvantage of direct model fits, is that features not included in the model are not tested for, and might drive the fit into a wrong region of parameter space when occurring in the light curve and when the model is forced to account for them (e.g. interpreting a plateau stage as regular decay with low $p$ value). The open source nature of the software should help facilitate the addition of additional dynamical models and radiative processes. This can be done both on the level of the the dynamical templates (e.g. by including a more complex evolution of the shock-accelerated electrons, \citealt{SironiGiannios2013}), and on the level of the spectral templates (e.g. shifting the cooling break directly, \citealt{Guidorzi2014}).

\section*{Acknowledgements}

I wish to thank Re'em Sari for helpful discussion. This work was supported by a post-doctoral Fellowship grant from the Alexander von Humboldt Foundation and by the DFG Cluster of Excellence "Origin and Structure of the Universe". The simulations have been carried out in part on the computing facilities of the Computational Center for Particle and Astrophysics (C2PAP).\\


\bibliographystyle{elsarticle-harv}  

\bibliography{proceedings_Rome_bibliography}

\begin{thebibliography}{142}
\expandafter\ifx\csname natexlab\endcsname\relax\def\natexlab#1{#1}\fi
\expandafter\ifx\csname url\endcsname\relax
  \def\url#1{\texttt{#1}}\fi
\expandafter\ifx\csname urlprefix\endcsname\relax\def\urlprefix{URL }\fi

\bibitem[{{Beloborodov}(2002)}]{Beloborodov2002}
{Beloborodov}, A.~M., Feb. 2002. {Radiation Front Sweeping the Ambient Medium
  of Gamma-Ray Bursts}. \apj 565, 808--828.

\bibitem[{{Berger} et~al.(2004){Berger}, {Kulkarni}, and {Frail}}]{Berger2004}
{Berger}, E., {Kulkarni}, S.~R., {Frail}, D.~A., Sep. 2004. {The
  Nonrelativistic Evolution of GRBs 980703 and 970508: Beaming-independent
  Calorimetry}. \apj 612, 966--973.

\bibitem[{{Berger} et~al.(2003)}]{Berger2003}
{Berger}, E., et~al., Nov. 2003. {A common origin for cosmic explosions
  inferred from calorimetry of GRB030329}. \nat 426, 154--157.

\bibitem[{{Beuermann} et~al.(1999)}]{Beuermann1999}
{Beuermann}, K., et~al., Dec. 1999. {VLT observations of GRB 990510 and its
  environment}. \aap 352, L26--L30.

\bibitem[{{Bhattacharya}(2001)}]{Bhattacharya2001}
{Bhattacharya}, D., Jun. 2001. {Flat Spectrum Gamma Ray Burst Afterglows}.
  Bulletin of the Astronomical Society of India 29, 107--114.

\bibitem[{{Blandford} and {McKee}(1976)}]{Blandford1976}
{Blandford}, R.~D., {McKee}, C.~F., Aug. 1976. {Fluid dynamics of relativistic
  blast waves}. Physics of Fluids 19, 1130--1138.

\bibitem[{{Bloom} et~al.(2003){Bloom}, {Frail}, and {Kulkarni}}]{Bloom2003}
{Bloom}, J.~S., {Frail}, D.~A., {Kulkarni}, S.~R., Sep. 2003. {Gamma-Ray Burst
  Energetics and the Gamma-Ray Burst Hubble Diagram: Promises and Limitations}.
  \apj 594, 674--683.

\bibitem[{{Buchner} et~al.(2014)}]{Buchner2014}
{Buchner}, J., et~al., Apr. 2014. {X-ray spectral modelling of the AGN
  obscuring region in the CDFS: Bayesian model selection and catalogue}. \aap
  564, A125.

\bibitem[{{Cannizzo} et~al.(2004){Cannizzo}, {Gehrels}, and
  {Vishniac}}]{CannizzoGehrelsVishniac2004}
{Cannizzo}, J.~K., {Gehrels}, N., {Vishniac}, E.~T., Jan. 2004. {A Numerical
  Gamma-Ray Burst Simulation Using Three-Dimensional Relativistic
  Hydrodynamics: The Transition from Spherical to Jetlike Expansion}. \apj 601,
  380--390.

\bibitem[{{Cenko} et~al.(2010)}]{Cenko2010}
{Cenko}, S.~B., et~al., Mar. 2010. {The Collimation and Energetics of the
  Brightest Swift Gamma-ray Bursts}. \apj 711, 641--654.

\bibitem[{{Cheng} et~al.(2001){Cheng}, {Huang}, and {Lu}}]{ChengHuangLu2001}
{Cheng}, K.~S., {Huang}, Y.~F., {Lu}, T., Aug. 2001. {Gamma-ray bursts:
  afterglows from cylindrical jets}. \mnras 325, 599--606.

\bibitem[{{Costa} et~al.(1997)}]{Costa1997}
{Costa}, E., et~al., Jun. 1997. {Discovery of an X-ray afterglow associated
  with the {$\gamma$}-ray burst of 28 February 1997}. \nat 387, 783--785.

\bibitem[{{Curran} et~al.(2009){Curran}, {Starling}, {van der Horst}, and
  {Wijers}}]{Curran2009}
{Curran}, P.~A., {Starling}, R.~L.~C., {van der Horst}, A.~J., {Wijers},
  R.~A.~M.~J., May 2009. {Testing the blast wave model with Swift GRBs}. \mnras
  395, 580--592.

\bibitem[{{Dai} and {Cheng}(2001)}]{DaiCheng2001}
{Dai}, Z.~G., {Cheng}, K.~S., Sep. 2001. {Afterglow Emission from Highly
  Collimated Jets with Flat Electron Spectra: Application to the GRB 010222
  Case?} \apjl 558, L109--L112.

\bibitem[{{Dai} and {Lu}(1998)}]{DaiLu1998}
{Dai}, Z.~G., {Lu}, T., May 1998. {Gamma-ray burst afterglows and evolution of
  postburst fireballs with energy injection from strongly magnetic millisecond
  pulsars}. \aap 333, L87--L90.

\bibitem[{{Dainotti} et~al.(2008){Dainotti}, {Cardone}, and
  {Capozziello}}]{Dainotti2008}
{Dainotti}, M.~G., {Cardone}, V.~F., {Capozziello}, S., Nov. 2008. {A
  time-luminosity correlation for {$\gamma$}-ray bursts in the X-rays}. \mnras
  391, L79--L83.

\bibitem[{{De Colle} et~al.(2012{\natexlab{a}}){De Colle}, {Granot},
  {L{\'o}pez-C{\'a}mara}, and {Ramirez-Ruiz}}]{DeColle2012Mezcal}
{De Colle}, F., {Granot}, J., {L{\'o}pez-C{\'a}mara}, D., {Ramirez-Ruiz}, E.,
  Feb. 2012{\natexlab{a}}. {Gamma-Ray Burst Dynamics and Afterglow Radiation
  from Adaptive Mesh Refinement, Special Relativistic Hydrodynamic
  Simulations}. \apj 746, 122.

\bibitem[{{De Colle} et~al.(2012{\natexlab{b}}){De Colle}, {Ramirez-Ruiz},
  {Granot}, and {Lopez-Camara}}]{DeColle2012}
{De Colle}, F., {Ramirez-Ruiz}, E., {Granot}, J., {Lopez-Camara}, D., May
  2012{\natexlab{b}}. {Simulations of Gamma-Ray Burst Jets in a Stratified
  External Medium: Dynamics, Afterglow Light Curves, Jet Breaks, and Radio
  Calorimetry}. \apj 751, 57.

\bibitem[{{De Pasquale} et~al.(2015)}]{DePasquale2015}
{De Pasquale}, M., et~al., Dec. 2015. {The optical rebrightening of GRB100814A:
  an interplay of forward and reverse shocks?} MNRAS Accepted. arXiv:1312.1648.

\bibitem[{{Downes} et~al.(2002){Downes}, {Duffy}, and
  {Komissarov}}]{Downes2002}
{Downes}, T.~P., {Duffy}, P., {Komissarov}, S.~S., May 2002. {Relativistic
  blast waves and synchrotron emission}. \mnras 332, 144--154.

\bibitem[{{Drenkhahn}(2002)}]{Drenkhahn2002}
{Drenkhahn}, G., May 2002. {Acceleration of GRB outflows by Poynting flux
  dissipation}. \aap 387, 714--724.

\bibitem[{{Duffell} and {MacFadyen}(2013)}]{DuffellMacFadyen2013}
{Duffell}, P.~C., {MacFadyen}, A.~I., Oct. 2013. {Rayleigh-Taylor Instability
  in a Relativistic Fireball on a Moving Computational Grid}. \apj 775, 87.

\bibitem[{{Duffell} and {MacFadyen}(2014)}]{DuffellMacFadyen2014}
{Duffell}, P.~C., {MacFadyen}, A.~I., Jul. 2014. {From Engine to Afterglow:
  Collapsars Naturally Produce Top-Heavy Jets and Early-Time Plateaus in Gamma
  Ray Burst Afterglows}. ArXiv:1407.8250.

\bibitem[{{Eichler} and {Granot}(2006)}]{EichlerGranot2006}
{Eichler}, D., {Granot}, J., Apr. 2006. {The Case for Anisotropic Afterglow
  Efficiency within Gamma-Ray Burst Jets}. \apjl 641, L5--L8.

\bibitem[{{Eichler} et~al.(1989){Eichler}, {Livio}, {Piran}, and
  {Schramm}}]{Eichler1989}
{Eichler}, D., {Livio}, M., {Piran}, T., {Schramm}, D.~N., Jul. 1989.
  {Nucleosynthesis, neutrino bursts and gamma-rays from coalescing neutron
  stars}. \nat 340, 126--128.

\bibitem[{{Eichler} and {Waxman}(2005)}]{EichlerWaxman2005}
{Eichler}, D., {Waxman}, E., Jul. 2005. {The Efficiency of Electron
  Acceleration in Collisionless Shocks and Gamma-Ray Burst Energetics}. \apj
  627, 861--867.

\bibitem[{{Eldridge}(2007)}]{Eldridge2007}
{Eldridge}, J.~J., May 2007. {Asymmetric Wolf-Rayet winds: implications for
  gamma-ray burst afterglows}. \mnras 377, L29--L33.

\bibitem[{{Eldridge} et~al.(2006){Eldridge}, {Genet}, {Daigne}, and
  {Mochkovitch}}]{Eldridge2006}
{Eldridge}, J.~J., {Genet}, F., {Daigne}, F., {Mochkovitch}, R., Mar. 2006.
  {The circumstellar environment of Wolf-Rayet stars and gamma-ray burst
  afterglows}. \mnras 367, 186--200.

\bibitem[{{Evans} et~al.(2009)}]{Evans2009}
{Evans}, P.~A., et~al., Aug. 2009. {Methods and results of an automatic
  analysis of a complete sample of Swift-XRT observations of GRBs}. \mnras 397,
  1177--1201.

\bibitem[{{Evans} et~al.(2014)}]{Evans2014}
{Evans}, P.~A., et~al., Oct. 2014. {GRB 130925A: an ultralong gamma ray burst
  with a dust-echo afterglow, and implications for the origin of the ultralong
  GRBs}. \mnras 444, 250--267.

\bibitem[{{Feroz} et~al.(2009){Feroz}, {Hobson}, and {Bridges}}]{Feroz2009}
{Feroz}, F., {Hobson}, M.~P., {Bridges}, M., Oct. 2009. {MULTINEST: an
  efficient and robust Bayesian inference tool for cosmology and particle
  physics}. \mnras 398, 1601--1614.

\bibitem[{{Filgas} et~al.(2011)}]{Filgas2011}
{Filgas}, R., et~al., Nov. 2011. {GRB 091127: The cooling break race on
  magnetic fuel}. \aap 535, A57.

\bibitem[{{Foreman-Mackey} et~al.(2013){Foreman-Mackey}, {Hogg}, {Lang}, and
  {Goodman}}]{Foreman-Mackey2013}
{Foreman-Mackey}, D., {Hogg}, D.~W., {Lang}, D., {Goodman}, J., Mar. 2013.
  {emcee: The MCMC Hammer}. \pasp 125, 306--312.

\bibitem[{{Galama}(1998)}]{Galama1998}
{Galama}, T.~J.~o., Oct. 1998. {An unusual supernova in the error box of the
  {$\gamma$}-ray burst of 25 April 1998}. \nat 395, 670--672.

\bibitem[{{Gao} et~al.(2013){Gao}, {Lei}, {Zou}, {Wu}, and {Zhang}}]{Gao2013}
{Gao}, H., {Lei}, W.-H., {Zou}, Y.-C., {Wu}, X.-F., {Zhang}, B., Dec. 2013. {A
  complete reference of the analytical synchrotron external shock models of
  gamma-ray bursts}. \nar 57, 141--190.

\bibitem[{{Gat} et~al.(2013){Gat}, {van Eerten}, and
  {MacFadyen}}]{GatvanEertenMacFadyen2013}
{Gat}, I., {van Eerten}, H., {MacFadyen}, A., Aug. 2013. {No Flares from
  Gamma-Ray Burst Afterglow Blast Waves Encountering Sudden Circumburst Density
  Change}. \apj 773, 2.

\bibitem[{{Gehrels} et~al.(2004)}]{Gehrels2004}
{Gehrels}, N., et~al., Aug. 2004. {The Swift Gamma-Ray Burst Mission}. \apj
  611, 1005--1020.

\bibitem[{{Geng} et~al.(2014){Geng}, {Wu}, {Li}, {Huang}, and {Dai}}]{Geng2014}
{Geng}, J.~J., {Wu}, X.~F., {Li}, L., {Huang}, Y.~F., {Dai}, Z.~G., Sep. 2014.
  {Revisiting the Emission from Relativistic Blast Waves in a Density-jump
  Medium}. \apj 792, 31.

\bibitem[{{Ghisellini} et~al.(2010){Ghisellini}, {Ghirlanda}, {Nava}, and
  {Celotti}}]{Ghisellini2010}
{Ghisellini}, G., {Ghirlanda}, G., {Nava}, L., {Celotti}, A., Apr. 2010. {GeV
  emission from gamma-ray bursts: a radiative fireball?} \mnras 403, 926--937.

\bibitem[{{Goodman} and {Weare}(2010)}]{GoodmanWeare2010}
{Goodman}, J., {Weare}, J., Jan. 2010. {Ensemble samplers with affine
  invariance}. Comm. App. Math. Comp. Sci. 5, 65--80.

\bibitem[{{Granot}(2007)}]{Granot2007}
{Granot}, J., Mar. 2007. {The Structure and Dynamics of GRB Jets}. In: Revista
  Mexicana de Astronomia y Astrofisica, vol. 27. Vol.~27 of Revista Mexicana de
  Astronomia y Astrofisica, vol. 27. pp. 140--165.

\bibitem[{{Granot} et~al.(2006){Granot}, {K{\"o}nigl}, and
  {Piran}}]{Granot2006}
{Granot}, J., {K{\"o}nigl}, A., {Piran}, T., Aug. 2006. {Implications of the
  early X-ray afterglow light curves of Swift gamma-ray bursts}. \mnras 370,
  1946--1960.

\bibitem[{{Granot} and {Piran}(2012)}]{GranotPiran2012}
{Granot}, J., {Piran}, T., Mar. 2012. {On the lateral expansion of gamma-ray
  burst jets}. \mnras 421, 570--587.

\bibitem[{{Granot} et~al.(1999){Granot}, {Piran}, and
  {Sari}}]{GranotPiranSari1999}
{Granot}, J., {Piran}, T., {Sari}, R., Dec. 1999. {Synchrotron Self-Absorption
  in Gamma-Ray Burst Afterglow}. \apj 527, 236--246.

\bibitem[{{Granot} and {Sari}(2002)}]{GranotSari2002}
{Granot}, J., {Sari}, R., Apr. 2002. {The Shape of Spectral Breaks in Gamma-Ray
  Burst Afterglows}. \apj 568, 820--829.

\bibitem[{{Greiner} et~al.(2011)}]{Greiner2011}
{Greiner}, J., et~al., Feb. 2011. {The nature of ``dark'' gamma-ray bursts}.
  \aap 526, A30.

\bibitem[{{Gruzinov} and {Waxman}(1999)}]{GruzinovWaxman1999}
{Gruzinov}, A., {Waxman}, E., Feb. 1999. {Gamma-Ray Burst Afterglow:
  Polarization and Analytic Light Curves}. \apj 511, 852--861.

\bibitem[{{Guidorzi} et~al.(2014)}]{Guidorzi2014}
{Guidorzi}, C., et~al., Feb. 2014. {New constraints on gamma-ray burst jet
  geometry and relativistic shock physics}. \mnras 438, 752--767.

\bibitem[{{Harrison} et~al.(1999)}]{Harrison1999}
{Harrison}, F.~A., et~al., Oct. 1999. {Optical and Radio Observations of the
  Afterglow from GRB 990510: Evidence for a Jet}. \apjl 523, L121--L124.

\bibitem[{{Huang} et~al.(1999){Huang}, {Dai}, and {Lu}}]{HuangDaiLu1999}
{Huang}, Y.~F., {Dai}, Z.~G., {Lu}, T., Oct. 1999. {A generic dynamical model
  of gamma-ray burst remnants}. \mnras 309, 513--516.

\bibitem[{{Huang} et~al.(2000){Huang}, {Gou}, {Dai}, and {Lu}}]{Huang2000}
{Huang}, Y.~F., {Gou}, L.~J., {Dai}, Z.~G., {Lu}, T., Nov. 2000. {Overall
  Evolution of Jetted Gamma-Ray Burst Ejecta}. \apj 543, 90--96.

\bibitem[{{Kobayashi} et~al.(1999){Kobayashi}, {Piran}, and
  {Sari}}]{Kobayashi1999}
{Kobayashi}, S., {Piran}, T., {Sari}, R., Mar. 1999. {Hydrodynamics of a
  Relativistic Fireball: The Complete Evolution}. \apj 513, 669--678.

\bibitem[{{Kouveliotou} et~al.(1993){Kouveliotou}, {Meegan}, {Fishman}, {Bhat},
  {Briggs}, {Koshut}, {Paciesas}, and {Pendleton}}]{Kouveliotou1993}
{Kouveliotou}, C., {Meegan}, C.~A., {Fishman}, G.~J., {Bhat}, N.~P., {Briggs},
  M.~S., {Koshut}, T.~M., {Paciesas}, W.~S., {Pendleton}, G.~N., Aug. 1993.
  {Identification of two classes of gamma-ray bursts}. \apjl 413, L101--L104.

\bibitem[{{Kumar} and {Panaitescu}(2000)}]{KumarPanaitescu2000}
{Kumar}, P., {Panaitescu}, A., Sep. 2000. {Steepening of Afterglow Decay for
  Jets Interacting with Stratified Media}. \apjl 541, L9--L12.

\bibitem[{{Landau} and {Lifshitz}(1959)}]{LandauLifshitz1959}
{Landau}, L.~D., {Lifshitz}, E.~M., 1959. {Fluid mechanics}.

\bibitem[{{Laskar} et~al.(2013)}]{Laskar2013}
{Laskar}, T., et~al., Oct. 2013. {A Reverse Shock in GRB 130427A}. \apj 776,
  119.

\bibitem[{{Leventis} et~al.(2013){Leventis}, {van der Horst}, {van Eerten}, and
  {Wijers}}]{Leventis2013}
{Leventis}, K., {van der Horst}, A.~J., {van Eerten}, H.~J., {Wijers},
  R.~A.~M.~J., May 2013. {Applying an accurate spherical model to gamma-ray
  burst afterglow observations}. \mnras 431, 1026--1038.

\bibitem[{{Leventis} et~al.(2012){Leventis}, {van Eerten}, {Meliani}, and
  {Wijers}}]{Leventis2012}
{Leventis}, K., {van Eerten}, H.~J., {Meliani}, Z., {Wijers}, R.~A.~M.~J., Dec.
  2012. {Practical flux prescriptions for gamma-ray burst afterglows, from
  early to late times}. \mnras 427, 1329--1343.

\bibitem[{{Leventis} et~al.(2014){Leventis}, {Wijers}, and {van der
  Horst}}]{Leventis2014}
{Leventis}, K., {Wijers}, R.~A.~M.~J., {van der Horst}, A.~J., Jan. 2014. {The
  plateau phase of gamma-ray burst afterglows in the thick-shell scenario}.
  \mnras 437, 2448--2460.

\bibitem[{{Li} et~al.(2012)}]{Li2012}
{Li}, L., et~al., Oct. 2012. {A Comprehensive Study of Gamma-Ray Burst Optical
  Emission. I. Flares and Early Shallow-decay Component}. \apj 758, 27.

\bibitem[{{Li} et~al.(2015)}]{Li2015}
{Li}, L., et~al., Mar. 2015. {A Correlated Study of Optical and X-ray
  Afterglows of GRBs}. ApJ accepted. ArXiv e-prints: 1503.00976.

\bibitem[{{Lyutikov}(2006)}]{lyutikov2006}
{Lyutikov}, M., Jul. 2006. {The electromagnetic model of gamma-ray bursts}. New
  Journal of Physics 8, 119.

\bibitem[{{MacFadyen} and {Woosley}(1999)}]{MacFadyenWoosley1999}
{MacFadyen}, A.~I., {Woosley}, S.~E., Oct. 1999. {Collapsars: Gamma-Ray Bursts
  and Explosions in ``Failed Supernovae''}. \apj 524, 262--289.

\bibitem[{{Meliani} et~al.(2007){Meliani}, {Keppens}, {Casse}, and
  {Giannios}}]{Meliani2007}
{Meliani}, Z., {Keppens}, R., {Casse}, F., {Giannios}, D., Apr. 2007. {AMRVAC
  and relativistic hydrodynamic simulations for gamma-ray burst afterglow
  phases}. \mnras 376, 1189--1200.

\bibitem[{{Mesler} et~al.(2012){Mesler}, {Whalen}, {Lloyd-Ronning}, {Fryer},
  and {Pihlstr{\"o}m}}]{Mesler2012}
{Mesler}, R.~A., {Whalen}, D.~J., {Lloyd-Ronning}, N.~M., {Fryer}, C.~L.,
  {Pihlstr{\"o}m}, Y.~M., Oct. 2012. {Gamma-Ray Bursts in Circumstellar Shells:
  A Possible Explanation for Flares}. \apj 757, 117.

\bibitem[{{Mesler} et~al.(2014){Mesler}, {Whalen}, {Smidt}, {Fryer},
  {Lloyd-Ronning}, and {Pihlstr{\"o}m}}]{Mesler2014}
{Mesler}, R.~A., {Whalen}, D.~J., {Smidt}, J., {Fryer}, C.~L., {Lloyd-Ronning},
  N.~M., {Pihlstr{\"o}m}, Y.~M., May 2014. {The First Gamma-Ray Bursts in the
  Universe}. \apj 787, 91.

\bibitem[{{M{\'e}sz{\'a}ros} and {Rees}(1997)}]{MeszarosRees1997}
{M{\'e}sz{\'a}ros}, P., {Rees}, M.~J., Feb. 1997. {Optical and Long-Wavelength
  Afterglow from Gamma-Ray Bursts}. \apj 476, 232--237.

\bibitem[{{M{\'e}sz{\'a}ros} and {Rees}(2000)}]{MeszarosRees2000}
{M{\'e}sz{\'a}ros}, P., {Rees}, M.~J., Feb. 2000. {Steep Slopes and Preferred
  Breaks in Gamma-Ray Burst Spectra: The Role of Photospheres and
  Comptonization}. \apj 530, 292--298.

\bibitem[{{Mignone} et~al.(2005){Mignone}, {Plewa}, and {Bodo}}]{Mignone2005}
{Mignone}, A., {Plewa}, T., {Bodo}, G., Sep. 2005. {The Piecewise Parabolic
  Method for Multidimensional Relativistic Fluid Dynamics}. \apjs 160,
  199--219.

\bibitem[{{Mimica} et~al.(2009{\natexlab{a}}){Mimica}, {Aloy}, {Agudo},
  {Mart{\'{\i}}}, {G{\'o}mez}, and {Miralles}}]{Mimica2009}
{Mimica}, P., {Aloy}, M.-A., {Agudo}, I., {Mart{\'{\i}}}, J.~M., {G{\'o}mez},
  J.~L., {Miralles}, J.~A., May 2009{\natexlab{a}}. {Spectral Evolution of
  Superluminal Components in Parsec-Scale Jets}. \apj 696, 1142--1163.

\bibitem[{{Mimica} and {Giannios}(2011)}]{MimicaGiannios2011}
{Mimica}, P., {Giannios}, D., Nov. 2011. {Gamma-ray burst afterglow light
  curves from realistic density profiles}. \mnras 418, 583--590.

\bibitem[{{Mimica} et~al.(2009{\natexlab{b}}){Mimica}, {Giannios}, and
  {Aloy}}]{MimicaGianniosAloy2009}
{Mimica}, P., {Giannios}, D., {Aloy}, M.~A., Feb. 2009{\natexlab{b}}.
  {Deceleration of arbitrarily magnetized GRB ejecta: the complete evolution}.
  \aap 494, 879--890.

\bibitem[{{Morsony} et~al.(2007){Morsony}, {Lazzati}, and
  {Begelman}}]{MorsonyLazzatiBegelman2007}
{Morsony}, B.~J., {Lazzati}, D., {Begelman}, M.~C., Aug. 2007. {Temporal and
  Angular Properties of Gamma-Ray Burst Jets Emerging from Massive Stars}. \apj
  665, 569--598.

\bibitem[{{Nakar} and {Granot}(2007)}]{NakarGranot2007}
{Nakar}, E., {Granot}, J., Oct. 2007. {Smooth light curves from a bumpy ride:
  relativistic blast wave encounters a density jump}. \mnras 380, 1744--1760.

\bibitem[{{Nava} et~al.(2013){Nava}, {Sironi}, {Ghisellini}, {Celotti}, and
  {Ghirlanda}}]{Nava2013}
{Nava}, L., {Sironi}, L., {Ghisellini}, G., {Celotti}, A., {Ghirlanda}, G.,
  Aug. 2013. {Afterglow emission in gamma-ray bursts - I. Pair-enriched ambient
  medium and radiative blast waves}. \mnras 433, 2107--2121.

\bibitem[{{Nousek} et~al.(2006)}]{Nousek2006}
{Nousek}, J.~A., et~al., May 2006. {Evidence for a Canonical Gamma-Ray Burst
  Afterglow Light Curve in the Swift XRT Data}. \apj 642, 389--400.

\bibitem[{{Oren} et~al.(2004){Oren}, {Nakar}, and {Piran}}]{Oren2004}
{Oren}, Y., {Nakar}, E., {Piran}, T., Oct. 2004. {The apparent size of
  gamma-ray burst afterglows as a test of the fireball model}. \mnras 353,
  L35--L40.

\bibitem[{{Paczynski}(1991)}]{Paczynski1991}
{Paczynski}, B., 1991. {Cosmological gamma-ray bursts}. \actaa 41, 257--267.

\bibitem[{{Paczynski}(1998)}]{Paczynski1998}
{Paczynski}, B., Feb. 1998. {Are Gamma-Ray Bursts in Star-Forming Regions?}
  \apjl 494, L45+.

\bibitem[{{Panaitescu} and {Kumar}(2000)}]{PanaitescuKumar2000}
{Panaitescu}, A., {Kumar}, P., Nov. 2000. {Analytic Light Curves of Gamma-Ray
  Burst Afterglows: Homogeneous versus Wind External Media}. \apj 543, 66--76.

\bibitem[{{Panaitescu} and {Kumar}(2001)}]{PanaitescuKumar2001}
{Panaitescu}, A., {Kumar}, P., Oct. 2001. {Fundamental Physical Parameters of
  Collimated Gamma-Ray Burst Afterglows}. \apjl 560, L49--L53.

\bibitem[{{Panaitescu} and {Kumar}(2002)}]{PanaitescuKumar2002}
{Panaitescu}, A., {Kumar}, P., Jun. 2002. {Properties of Relativistic Jets in
  Gamma-Ray Burst Afterglows}. \apj 571, 779--789.

\bibitem[{{Panaitescu} et~al.(1998){Panaitescu}, {Meszaros}, and
  {Rees}}]{Panaitescu1998}
{Panaitescu}, A., {Meszaros}, P., {Rees}, M.~J., Aug. 1998. {Multiwavelength
  Afterglows in Gamma-Ray Bursts: Refreshed Shock and Jet Effects}. \apj 503,
  314.

\bibitem[{{Panaitescu} and {Vestrand}(2011)}]{PanaitescuVestrand2011}
{Panaitescu}, A., {Vestrand}, W.~T., Jul. 2011. {Optical afterglows of
  gamma-ray bursts: peaks, plateaus and possibilities}. \mnras 414, 3537--3546.

\bibitem[{{Pe'er}(2012)}]{Peer2012}
{Pe'er}, A., Jun. 2012. {Dynamical Model of an Expanding Shell}. \apjl 752, L8.

\bibitem[{{Perley} et~al.(2014)}]{Perley2014}
{Perley}, D.~A., et~al., Jan. 2014. {The Afterglow of GRB 130427A from 1 to
  10$^{16}$ GHz}. \apj 781, 37.

\bibitem[{{Piran}(1999)}]{Piran1999}
{Piran}, T., Jun. 1999. {Gamma-ray bursts and the fireball model}. \physrep
  314, 575--667.

\bibitem[{{Racusin} et~al.(2009){Racusin}, {Liang}, {Burrows}, {Falcone},
  {Sakamoto}, {Zhang}, {Zhang}, {Evans}, and {Osborne}}]{Racusin2009}
{Racusin}, J.~L., {Liang}, E.~W., {Burrows}, D.~N., {Falcone}, A., {Sakamoto},
  T., {Zhang}, B.~B., {Zhang}, B., {Evans}, P., {Osborne}, J., Jun. 2009. {Jet
  Breaks and Energetics of Swift Gamma-Ray Burst X-Ray Afterglows}. \apj 698,
  43--74.

\bibitem[{{Racusin} et~al.(2011)}]{Racusin2011}
{Racusin}, J.~L., et~al., Sep. 2011. {Fermi and Swift Gamma-ray Burst Afterglow
  Population Studies}. \apj 738, 138.

\bibitem[{{Rees} and {Meszaros}(1992)}]{ReesMeszaros1992}
{Rees}, M.~J., {Meszaros}, P., Sep. 1992. {Relativistic fireballs - Energy
  conversion and time-scales}. \mnras 258, 41P--43P.

\bibitem[{{Rees} and {Meszaros}(1998)}]{ReesMeszaros1998}
{Rees}, M.~J., {Meszaros}, P., Mar. 1998. {Refreshed Shocks and Afterglow
  Longevity in Gamma-Ray Bursts}. \apjl 496, L1.

\bibitem[{{Rhoads}(1997)}]{Rhoads1997}
{Rhoads}, J.~E., Sep. 1997. {How to Tell a Jet from a Balloon: A Proposed Test
  for Beaming in Gamma-Ray Bursts}. \apjl 487, L1--L4.

\bibitem[{{Rhoads}(1999)}]{Rhoads1999}
{Rhoads}, J.~E., Nov. 1999. {The Dynamics and Light Curves of Beamed Gamma-Ray
  Burst Afterglows}. \apj 525, 737--749.

\bibitem[{{Rossi} et~al.(2002){Rossi}, {Lazzati}, and {Rees}}]{Rossi2002}
{Rossi}, E., {Lazzati}, D., {Rees}, M.~J., Jun. 2002. {Afterglow light curves,
  viewing angle and the jet structure of {$\gamma$}-ray bursts}. \mnras 332,
  945--950.

\bibitem[{{Ryan} et~al.(2015){Ryan}, {van Eerten}, {MacFadyen}, and
  {Zhang}}]{Ryan2015}
{Ryan}, G., {van Eerten}, H., {MacFadyen}, A., {Zhang}, B.-B., Jan. 2015.
  {Gamma-Ray Bursts are Observed Off-axis}. \apj 799, 3.

\bibitem[{{Rybicki} and {Lightman}(1979)}]{RybickiLightman1979}
{Rybicki}, G.~B., {Lightman}, A.~P., 1979. {Radiative processes in
  astrophysics}.

\bibitem[{{Sari} and {Esin}(2001)}]{SariEsin2001}
{Sari}, R., {Esin}, A.~A., Feb. 2001. {On the Synchrotron Self-Compton Emission
  from Relativistic Shocks and Its Implications for Gamma-Ray Burst
  Afterglows}. \apj 548, 787--799.

\bibitem[{{Sari} and {M{\'e}sz{\'a}ros}(2000)}]{SariMeszaros2000}
{Sari}, R., {M{\'e}sz{\'a}ros}, P., May 2000. {Impulsive and Varying Injection
  in Gamma-Ray Burst Afterglows}. \apjl 535, L33--L37.

\bibitem[{{Sari} et~al.(1996){Sari}, {Narayan}, and
  {Piran}}]{SariNarayanPiran1996}
{Sari}, R., {Narayan}, R., {Piran}, T., Dec. 1996. {Cooling Timescales and
  Temporal Structure of Gamma-Ray Bursts}. \apj 473, 204.

\bibitem[{{Sari} and {Piran}(1995)}]{SariPiran1995}
{Sari}, R., {Piran}, T., Dec. 1995. {Hydrodynamic Timescales and Temporal
  Structure of Gamma-Ray Bursts}. \apjl 455, L143.

\bibitem[{{Sari} et~al.(1999){Sari}, {Piran}, and
  {Halpern}}]{SariPiranHalpern1999}
{Sari}, R., {Piran}, T., {Halpern}, J.~P., Jul. 1999. {Jets in Gamma-Ray
  Bursts}. \apjl 519, L17--L20.

\bibitem[{{Sari} et~al.(1998){Sari}, {Piran}, and
  {Narayan}}]{SariPiranNarayan1998}
{Sari}, R., {Piran}, T., {Narayan}, R., Apr. 1998. {Spectra and Light Curves of
  Gamma-Ray Burst Afterglows}. \apjl 497, L17--L20.

\bibitem[{{Sedov}(1959)}]{Sedov1959}
{Sedov}, L.~I., 1959. {Similarity and Dimensional Methods in Mechanics}.
  {Similarity and Dimensional Methods in Mechanics, New York: Academic Press,
  1959}.

\bibitem[{{Shivvers} and {Berger}(2011)}]{Shivvers2011}
{Shivvers}, I., {Berger}, E., Jun. 2011. {A Beaming-independent Estimate of the
  Energy Distribution of Long Gamma-Ray Bursts: Initial Results and Future
  Prospects}. \apj 734, 58.

\bibitem[{{Sironi} and {Giannios}(2013)}]{SironiGiannios2013}
{Sironi}, L., {Giannios}, D., Dec. 2013. {A Late-time Flattening of Light
  Curves in Gamma-Ray Burst Afterglows}. \apj 778, 107.

\bibitem[{{Soderberg} et~al.(2006){Soderberg}, {Nakar}, {Berger}, and
  {Kulkarni}}]{Soderberg2006}
{Soderberg}, A.~M., {Nakar}, E., {Berger}, E., {Kulkarni}, S.~R., Feb. 2006.
  {Late-Time Radio Observations of 68 Type Ibc Supernovae: Strong Constraints
  on Off-Axis Gamma-Ray Bursts}. \apj 638, 930--937.

\bibitem[{{Taylor}(1950)}]{Taylor1950}
{Taylor}, G., Mar. 1950. {The Formation of a Blast Wave by a Very Intense
  Explosion. I. Theoretical Discussion}. Royal Society of London Proceedings
  Series A 201, 159--174.

\bibitem[{{Taylor} et~al.(2004){Taylor}, {Frail}, {Berger}, and
  {Kulkarni}}]{Taylor2004}
{Taylor}, G.~B., {Frail}, D.~A., {Berger}, E., {Kulkarni}, S.~R., Jul. 2004.
  {The Angular Size and Proper Motion of the Afterglow of GRB 030329}. \apjl
  609, L1--L4.

\bibitem[{{Thompson}(1994)}]{Thompson1994}
{Thompson}, C., Oct. 1994. {A Model of Gamma-Ray Bursts}. \mnras 270, 480.

\bibitem[{{Thompson} and {Madau}(2000)}]{ThompsonMadau2000}
{Thompson}, C., {Madau}, P., Jul. 2000. {Relativistic Winds from Compact
  Gamma-Ray Sources. II. Pair Loading and Radiative Acceleration in Gamma-Ray
  Bursts}. \apj 538, 105--114.

\bibitem[{{Uhm}(2011)}]{Uhm2011}
{Uhm}, Z.~L., Jun. 2011. {A Semi-analytic Formulation for Relativistic Blast
  Waves with a Long-lived Reverse Shock}. \apj 733, 86.

\bibitem[{{Uhm} and {Zhang}(2014)}]{UhmZhang2014}
{Uhm}, Z.~L., {Zhang}, B., Jul. 2014. {Dynamics and Afterglow Light Curves of
  Gamma-Ray Burst Blast Waves Encountering a Density Bump or Void}. \apj 789,
  39.

\bibitem[{{Uhm} et~al.(2012){Uhm}, {Zhang}, {Hasco{\"e}t}, {Daigne},
  {Mochkovitch}, and {Park}}]{Uhm2012}
{Uhm}, Z.~L., {Zhang}, B., {Hasco{\"e}t}, R., {Daigne}, F., {Mochkovitch}, R.,
  {Park}, I.~H., Dec. 2012. {Dynamics and Afterglow Light Curves of Gamma-Ray
  Burst Blast Waves with a Long-lived Reverse Shock}. \apj 761, 147.

\bibitem[{{Urata} et~al.(2014)}]{Urata2014}
{Urata}, Y., et~al., Jul. 2014. {Synchrotron Self-inverse Compton Radiation
  from Reverse Shock on GRB 120326A}. \apj 789, 146.

\bibitem[{{Usov}(1992)}]{Usov1992}
{Usov}, V.~V., Jun. 1992. {Millisecond pulsars with extremely strong magnetic
  fields as a cosmological source of gamma-ray bursts}. \nat 357, 472--474.

\bibitem[{{van Eerten}(2013)}]{vanEerten2013}
{van Eerten}, H., Sep. 2013. {Gamma-ray burst afterglow theory}. eConf
  Proceedings 7th Huntsville Gamma-Ray Burst Symposium, GRB 2013, 24.

\bibitem[{{van Eerten}(2014{\natexlab{a}})}]{vanEerten2014injection}
{van Eerten}, H., Aug. 2014{\natexlab{a}}. {Self-similar relativistic blast
  waves with energy injection}. \mnras 442, 3495--3510.

\bibitem[{{van Eerten} and {MacFadyen}(2013)}]{vanEertenMacFadyen2013}
{van Eerten}, H., {MacFadyen}, A., Apr. 2013. {Gamma-Ray Burst Afterglow Light
  Curves from a Lorentz-boosted Simulation Frame and the Shape of the Jet
  Break}. \apj 767, 141.

\bibitem[{{van Eerten} et~al.(2012){van Eerten}, {van der Horst}, and
  {MacFadyen}}]{vanEertenvanderHorstMacFadyen2012}
{van Eerten}, H., {van der Horst}, A., {MacFadyen}, A., Apr. 2012. {Gamma-Ray
  Burst Afterglow Broadband Fitting Based Directly on Hydrodynamics
  Simulations}. \apj 749, 44.

\bibitem[{{van Eerten} et~al.(2010{\natexlab{a}}){van Eerten}, {Zhang}, and
  {MacFadyen}}]{vanEertenZhangMacFadyen2010}
{van Eerten}, H., {Zhang}, W., {MacFadyen}, A., Oct. 2010{\natexlab{a}}.
  {Off-axis Gamma-ray Burst Afterglow Modeling Based on a Two-dimensional
  Axisymmetric Hydrodynamics Simulation}. \apj 722, 235--247.

\bibitem[{{van Eerten}(2014{\natexlab{b}})}]{vanEerten2014}
{van Eerten}, H.~J., Dec. 2014{\natexlab{b}}. {Gamma-ray burst afterglow
  plateau break time-luminosity correlations favour thick shell models over
  thin shell models}. \mnras 445, 2414--2423.

\bibitem[{{van Eerten} et~al.(2010{\natexlab{b}}){van Eerten}, {Leventis},
  {Meliani}, {Wijers}, and {Keppens}}]{vanEerten2010transrelativistic}
{van Eerten}, H.~J., {Leventis}, K., {Meliani}, Z., {Wijers}, R.~A.~M.~J.,
  {Keppens}, R., Mar. 2010{\natexlab{b}}. {Gamma-ray burst afterglows from
  transrelativistic blast wave simulations}. \mnras 403, 300--316.

\bibitem[{{van Eerten} and
  {MacFadyen}(2012{\natexlab{a}})}]{vanEertenMacFadyen2012scaling}
{van Eerten}, H.~J., {MacFadyen}, A.~I., Mar. 2012{\natexlab{a}}. {Gamma-Ray
  Burst Afterglow Scaling Relations for the Full Blast Wave Evolution}. \apjl
  747, L30.

\bibitem[{{van Eerten} and
  {MacFadyen}(2012{\natexlab{b}})}]{vanEertenMacFadyen2012}
{van Eerten}, H.~J., {MacFadyen}, A.~I., Jun. 2012{\natexlab{b}}.
  {Observational Implications of Gamma-Ray Burst Afterglow Jet Simulations and
  Numerical Light Curve Calculations}. \apj 751, 155.

\bibitem[{{van Eerten} et~al.(2009){van Eerten}, {Meliani}, {Wijers}, and
  {Keppens}}]{vanEerten2009}
{van Eerten}, H.~J., {Meliani}, Z., {Wijers}, R.~A.~M.~J., {Keppens}, R., Sep.
  2009. {No visible optical variability from a relativistic blast wave
  encountering a wind termination shock}. \mnras 398, L63--L67.

\bibitem[{{van Eerten} et~al.(2011){van Eerten}, {Meliani}, {Wijers}, and
  {Keppens}}]{vanEerten2011chromaticbreak}
{van Eerten}, H.~J., {Meliani}, Z., {Wijers}, R.~A.~M.~J., {Keppens}, R., Jan.
  2011. {Jet simulations and gamma-ray burst afterglow jet breaks}. \mnras 410,
  2016--2024.

\bibitem[{{van Eerten} and {Wijers}(2009)}]{vanEertenWijers2009}
{van Eerten}, H.~J., {Wijers}, R.~A.~M.~J., Apr. 2009. {Gamma-ray burst
  afterglow scaling coefficients for general density profiles}. \mnras 394,
  2164--2174.

\bibitem[{{van Marle} and {Keppens}(2012)}]{vanMarleKeppens2012}
{van Marle}, A.~J., {Keppens}, R., Nov. 2012. {Multi-dimensional models of
  circumstellar shells around evolved massive stars}. \aap 547, A3.

\bibitem[{{van Marle} et~al.(2006){van Marle}, {Langer}, {Achterberg}, and
  {Garc{\'{\i}}a-Segura}}]{vanMarle2006}
{van Marle}, A.~J., {Langer}, N., {Achterberg}, A., {Garc{\'{\i}}a-Segura}, G.,
  Dec. 2006. {Forming a constant density medium close to long gamma-ray
  bursts}. \aap 460, 105--116.

\bibitem[{{van Marle} et~al.(2008){van Marle}, {Langer}, {Yoon}, and
  {Garc{\'{\i}}a-Segura}}]{vanMarle2008}
{van Marle}, A.~J., {Langer}, N., {Yoon}, S.-C., {Garc{\'{\i}}a-Segura}, G.,
  Feb. 2008. {The circumstellar medium around a rapidly rotating, chemically
  homogeneously evolving, possible gamma-ray burst progenitor}. \aap 478,
  769--778.

\bibitem[{{van Paradijs} et~al.(1997)}]{vanParadijs1997}
{van Paradijs}, J., et~al., Apr. 1997. {Transient optical emission from the
  error box of the {$\gamma$}-ray burst of 28 February 1997}. \nat 386,
  686--689.

\bibitem[{{Waxman} et~al.(1998){Waxman}, {Kulkarni}, and
  {Frail}}]{WaxmanKulkarniFrail1998}
{Waxman}, E., {Kulkarni}, S.~R., {Frail}, D.~A., Apr. 1998. {Implications of
  the Radio Afterglow from the Gamma-Ray Burst of 1997 May 8}. \apj 497,
  288--293.

\bibitem[{{Waxman} and {M{\'e}sz{\'a}ros}(2003)}]{WaxmanMeszaros2003}
{Waxman}, E., {M{\'e}sz{\'a}ros}, P., Feb. 2003. {Collapsar Uncorking and Jet
  Eruption in Gamma-Ray Bursts}. \apj 584, 390--398.

\bibitem[{{Wijers} et~al.(1997){Wijers}, {Rees}, and {Meszaros}}]{Wijers1997}
{Wijers}, R.~A.~M.~J., {Rees}, M.~J., {Meszaros}, P., Jul. 1997. {Shocked by
  GRB 970228: the afterglow of a cosmological fireball}. \mnras 288, L51--L56.

\bibitem[{{Woosley}(1993)}]{Woosley1993}
{Woosley}, S.~E., Mar. 1993. {Gamma-ray bursts from stellar mass accretion
  disks around black holes}. \apj 405, 273--277.

\bibitem[{{Wygoda} et~al.(2011){Wygoda}, {Waxman}, and {Frail}}]{Wygoda2011}
{Wygoda}, N., {Waxman}, E., {Frail}, D.~A., Sep. 2011. {Relativistic Jet
  Dynamics and Calorimetry of Gamma-ray Bursts}. \apjl 738, L23.

\bibitem[{{Yi} et~al.(2013){Yi}, {Wu}, and {Dai}}]{YiWuDai2013}
{Yi}, S.-X., {Wu}, X.-F., {Dai}, Z.-G., Oct. 2013. {Early Afterglows of
  Gamma-Ray Bursts in a Stratified Medium with a Power-law Density
  Distribution}. \apj 776, 120.

\bibitem[{{Yost} et~al.(2003){Yost}, {Harrison}, {Sari}, and
  {Frail}}]{Yost2003}
{Yost}, S.~A., {Harrison}, F.~A., {Sari}, R., {Frail}, D.~A., Nov. 2003. {A
  Study of the Afterglows of Four Gamma-Ray Bursts: Constraining the Explosion
  and Fireball Model}. \apj 597, 459--473.

\bibitem[{{Zhang} et~al.(2006){Zhang}, {Fan}, {Dyks}, {Kobayashi},
  {M{\'e}sz{\'a}ros}, {Burrows}, {Nousek}, and {Gehrels}}]{ZhangBing2006}
{Zhang}, B., {Fan}, Y.~Z., {Dyks}, J., {Kobayashi}, S., {M{\'e}sz{\'a}ros}, P.,
  {Burrows}, D.~N., {Nousek}, J.~A., {Gehrels}, N., May 2006. {Physical
  Processes Shaping Gamma-Ray Burst X-Ray Afterglow Light Curves: Theoretical
  Implications from the Swift X-Ray Telescope Observations}. \apj 642,
  354--370.

\bibitem[{{Zhang} and {M{\'e}sz{\'a}ros}(2001)}]{ZhangMeszaros2001}
{Zhang}, B., {M{\'e}sz{\'a}ros}, P., May 2001. {Gamma-Ray Burst Afterglow with
  Continuous Energy Injection: Signature of a Highly Magnetized Millisecond
  Pulsar}. \apjl 552, L35--L38.

\bibitem[{{Zhang} et~al.(2014){Zhang}, {van Eerten}, {Burrows}, {Ryan},
  {Evans}, {Racusin}, {Troja}, and {MacFadyen}}]{ZhangBinbin2015}
{Zhang}, B.-B., {van Eerten}, H., {Burrows}, D.~N., {Ryan}, G.~S., {Evans},
  P.~A., {Racusin}, J.~L., {Troja}, E., {MacFadyen}, A., May 2014. {An Analysis
  of Chandra Deep Follow-up GRBs: Implications for Off-Axis Jets}. ApJ
  accepted. ArXiv e-prints: 1405.4867.

\bibitem[{{Zhang} and {MacFadyen}(2009)}]{ZhangMacFadyen2009}
{Zhang}, W., {MacFadyen}, A., Jun. 2009. {The Dynamics and Afterglow Radiation
  of Gamma-Ray Bursts. I. Constant Density Medium}. \apj 698, 1261--1272.

\end{thebibliography}

\end{document}